\newcommand{\enableAFIVE}[1]{}
\newcommand{\enableAFOUR}[1]{#1}

\documentclass[a4paper,12pt]{mybook}

\newcommand{\erdos}{Erd{\H o}s-R{\'e}nyi}

\newcommand{\mycomment}[1]{}

\newcounter{firstbib}
\usepackage[utf8]{inputenc}
\usepackage[T1]{fontenc}
\usepackage{ngerman}
\usepackage{etoolbox}
\usepackage{amssymb}
\usepackage{multibib}
\usepackage{amsmath}
\usepackage{mathptmx} 
\usepackage{graphicx}
\usepackage{wrapfig}
\usepackage{parskip}
\usepackage{textcomp}
\usepackage{tabularx}
\usepackage{bbold}
\usepackage{numprint}
\usepackage{pdflscape}

\usepackage{url}

\usepackage[monochrome]{color}

\usepackage{url}
\usepackage{multirow}
\usepackage{algorithmic}
\usepackage{algorithm}
\usepackage{colortbl}
\usepackage{rotating}
\usepackage{subfigure}

\usepackage{algorithm}
\npdecimalsign{.} 

\usepackage[linktocpage]{hyperref}
\hypersetup{%
   colorlinks=true,
   citecolor=blue,
   urlcolor=blue,
   breaklinks=true,
   pdftitle={High Quality Graph Partitioning},
   pdfsubject={PhD thesis},
   pdfauthor={Christian Schulz},
   pdfproducer={}
}

\enableAFIVE{
\sloppy
        \hfuzz 1.5pt

        \hoffset=-1in
        \voffset=-1in
        \setlength{\textwidth}{116mm}
\setlength{\oddsidemargin}{14mm}
\setlength{\evensidemargin}{18mm}
        \setlength{\headheight}{20mm}
\setlength{\headsep}{4mm}
\setlength{\textheight}{174.8mm}
\setlength{\topmargin}{-6mm}

}
\enableAFOUR{
\sloppy
\hfuzz 1.5pt

\hoffset=-1in
\voffset=-1in
\setlength{\textwidth}{148mm}
\setlength{\oddsidemargin}{34mm}
\setlength{\evensidemargin}{28mm}
\setlength{\headheight}{20mm}
\setlength{\headsep}{4mm}
\setlength{\textheight}{215mm}
\setlength{\topmargin}{18mm}
}
\usepackage{myfancyheadings}
\graphicspath{{fig/}{plot/}}

{\bfseries}{\itshape}
{\bfseries}{\itshape}
{\bfseries}{\itshape}
{\bfseries}{\itshape}
{\bfseries}{\itshape}
{\bfseries}{\itshape}

\newcommand{\mycoauthor}[1]{{\leftskip=20pt \rightskip=20pt \textit{#1} \\ \par}}

\definecolor{dgreen}{rgb}{0,0.56,0} 
\definecolor{dmagenta}{rgb}{0.56,0,0.56} 
\definecolor{dcyan}{rgb}{0,0.56,0.56} 
\definecolor{dbrown}{rgb}{0.5,0.19,0} 
\definecolor{highlight}{rgb}{0,0,0}
\definecolor{lblue}{rgb}{0.9,0.9,1}

\newcommand{\floor}[1]{\left\lfloor #1\right\rfloor}

\newcommand{\set}[1]{\left\{ #1\right\}}
\newcommand{\gilt}{:}

\newcommand{\setGilt}[2]{\left\{ #1\gilt #2\right\}}

\newcommand{\realrange}[2]{\left[#1, #2\right]}

\newcommand{\unitrange}[2]{\realrange{0}{1}}

\newcommand{\Oh}[1]{\mathrm{O}\!\left( #1\right)}

\newcommand{\discussionsize}{\small}

\newcommand{\notiz}[1]{}

\newsavebox{\codeparam}
\newcounter{lineNumber}
{\end{disscodepos}}

\newcommand{\Is}{\mbox{\rm := }}
\newcommand{\ie}{i.e.\ }
\newdimen\endofsize\endofsize=0.5em

\newcommand{\etal}{et~al.\ }
\newcommand{\eg}{e.g.\ }

\usepackage{todonotes}

\usepackage[ngerman,english]{babel}
\begin{document}

\pagestyle{empty}
\setcounter{page}{1}
\begin{center}
\vspace*{10mm}
\LARGE
{\bf Scalable Graph Algorithms}\\[10mm]

\large
{\bf Zur Erlangung der Lehrbefähigung\\ im Fach Informatik\\[10mm]

angenommene\\[10mm]
                                 
\Large
Kumulative Habilitationsschrift\\[12mm]}

{von}\\[12mm]

\large
{Dr. rer. nat. Christian Schulz}\\[5mm]

\end{center}

\vfill
\begin{tabbing}
\large Einreichung: \hspace*{36mm}\=April 2019\\[3.5mm]
\large Mdl.~Prüfung: \hspace*{33mm}\=November 2019\\[3.5mm]
\end{tabbing}

\newpage
\cleardoublepage
\pagestyle{fancyplain}
\tableofcontents
\chapter*{List of Submitted Publications}
\addcontentsline{toc}{chapter}{List of Submitted Publications}
This cumulative habilitation thesis contains the following papers on graph algorithms.

\renewcommand\refname{}
\section*{Multilevel Algorithms}       
\begin{thenewbibliography}{100}
\bibitem{a}
Henning Meyerhenke, Peter Sanders and Christian Schulz.
\newblock {Partitioning Complex Networks via Size-constrained Clustering}.
\newblock In {\em Proceedings of the 13th Symposium on Experimental Algorithms (SEA)}, volume 8504 of Lecture Notes in Computer Science, pages 351--363. Springer, 2014.
\bibitem{a}
Yaroslav Akhremtsev, Peter Sanders and Christian Schulz.
\newblock {(Semi-)External Algorithms for Graph Partitioning and Clustering}.
\newblock In {\em Proceedings of the 17th Workshop on Algorithm Engineering and Experimentation (ALENEX)}, pages 33--43. SIAM,~2015.

\bibitem{a}
Henning Meyerhenke, Peter Sanders and Christian Schulz.
\newblock {Parallel Graph Partitioning for Complex Networks}.
\newblock In {\em 29th IEEE International Parallel and Distributed Processing Symposium (IPDPS)}, 2015.

\bibitem{a}
Henning Meyerhenke, Martin Nöllenburg and Christian Schulz.
\newblock {Drawing Large Graphs by Multilevel Maxent-Stress Optimization}.
\newblock In {\em Proceedings of the 23rd International Symposium on
Graph Drawing \& Network Visualization (GD)}, volume 9411 of Lecture Notes in Computer Science, pages 30--43. Springer, 2015.

\bibitem{a}
Sebastian Schlag, Vitali Henne, Tobias Heuer, Henning Meyerhenke, Peter Sanders and Christian Schulz.
\newblock {$k$-way Hypergraph Partitioning via $n$-Level Recursive Bisection}.
\newblock In {\em Proceedings of the 18th Workshop on Algorithm Engineering and Experimentation (ALENEX)}, pages 53-67. SIAM 2016.
\bibitem{a}
Peter Sanders and Christian Schulz.
\newblock {Advanced Multilevel Node Separator Algorithms}.
\newblock In {\em Proceedings of the 15th Symposium on Experimental Algorithms (SEA)}, volume 9685 of Lecture Notes in Computer Science, pages 294--309. Springer,~2016.

\bibitem{qapone}
Yaroslav Akhremtsev, Peter Sanders and Christian Schulz.
\newblock {High-Quality Shared-Memory Graph Partitioning}.
\newblock In {\em Proceedings of the 24th International European Conference on Parallel Computing (Euro-Par)}, volume 11014 of LNCS, pages 659--671, 2018.

\bibitem{a}
Christian Schulz and Jesper Larsson Träff.
\newblock {Better Process Mapping and Sparse Quadratic Assignment Problems}.
\newblock In {\em Proceedings of the 16th Symposium on Experimental Algorithms (SEA)}, volume 75 of LIPIcs, pages 4:1--4:15, 2017.

\bibitem{a}
Orlando Moreira, Merten Popp and Christian Schulz.
\newblock {Graph Partitioning with Acyclicity Constraints}.
\newblock In {\em Proceedings of the 16th Symposium on Experimental Algorithms (SEA)}, volume 75 of LIPIcs, pages 30:1--30:15, 2017.

\bibitem{a}
Henning Meyerhenke, Peter Sanders and Christian Schulz.
\newblock {Partitioning (Hierarchically Clustered) Complex Networks via Size-Constrained Graph Clustering}.
\newblock {\em ACM Journal of Heuristics}, Volume 22, Issue 5, pages 759--782, 2016.

\bibitem{a}
Henning Meyerhenke, Peter Sanders and Christian Schulz.
\newblock {Parallel Graph Partitioning for Complex Networks}.
\newblock {\em IEEE Transactions on Parallel and Distributed Systems}, Volume 28, Issue 9, pages 2625--2638, 2017.

\bibitem{jvdrawing}
Henning Meyerhenke, Martin Nöllenburg and Christian Schulz.
\newblock {Drawing Large Graphs by Multilevel Maxent-Stress Optimization}.
\newblock {\em IEEE Transactions on Visualization and Computer Graphics}, Volume 24, Issue 5, pages 1814--1827. 2018.

\setcounter{firstbib}{\value{enumiv}}
\end{thenewbibliography}
\section*{Practical Kernelization}       
\begin{thenewbibliography}{100}
\setcounter{enumiv}{\value{firstbib}}
\bibitem{a}
Sebastian Lamm, Peter Sanders, Christian Schulz, Darren Strash and Renato F.\@ Werneck.
\newblock {Finding Near-Optimal Independent Sets at Scale}.
\newblock In {\em Proceedings of the 18th Workshop on Algorithm Engineering and Experimentation (ALENEX)}, pages 138--150. SIAM~2016.

\bibitem{a}
Jakob Dahlum, Sebastian Lamm, Peter Sanders, Christian Schulz, Darren Strash and Renato F.\@ Werneck.
\newblock {Accelerating Local Search for the Maximum Independent Set Problem}.
\newblock In {\em Proceedings of the 15th Symposium on Experimental Algorithms (SEA)},  volume 9685 of Lecture Notes in Computer Science, pages 118--133. Springer, 2016.

\bibitem{a}
Monika Henzinger, Alexander Noe, Christian Schulz and Darren Strash.
\newblock {Practical Minimum Cut Algorithms}.
\newblock In {\em Proceedings of the 20th Workshop on Algorithm Engineering and Experimentation (ALENEX)}, pages 48--61. SIAM~2018.

\bibitem{a}
Demian Hespe, Christian Schulz and Darren Strash.
\newblock {Scalable Kernelization for Maximum Independent Sets}.
\newblock In {\em Proceedings of the 20th Workshop on Algorithm Engineering and Experimentation (ALENEX)}, pages 223--237, 2018.

\bibitem{qapone}
Monika Henzinger, Alexander Noe and Christian Schulz.
\newblock {Shared-Memory Exact Minimum Cuts}.
\newblock In {\em 33rd IEEE International Parallel and Distributed Processing Symposium (IPDPS)}, \emph{to appear}, 2019. 

\bibitem{qapone}
Sebastian Lamm, Christian Schulz, Darren Strash, Robert Williger and Huashuo Zhang.
\newblock {Exactly Solving the Maximum Weight Independent Set Problem on Large Real-World Graphs}.
\newblock In {\em Proceedings of the 21th Workshop on Algorithm Engineering and Experimentation (ALENEX)}, pages 144--158, SIAM, 2019. 

\bibitem{a}
Christian Schulz, Darren Strash, Sebastian Lamm, Peter Sanders and Renato F.\@ Werneck.
\newblock {Finding Near-Optimal Independent Sets at Scale}.
\newblock {\em ACM Journal of Heuristics}, Volume 23, Issue 4, pages 207--229, 2017.

\bibitem{qapone}
Monika Henzinger, Alexander Noe, Christian Schulz and Darren Strash.
\newblock {Practical Minimum Cut Algorithms}.
\newblock Invited to special issue of {\em ACM Journal of Experimental Algorithms (ACM JEA) for ALENEX 2018}, Volume 23, pages 1.9:1--1.8:22,~2018.

\setcounter{firstbib}{\value{enumiv}}
\end{thenewbibliography}
\section*{More Parallelization}       
\begin{thenewbibliography}{100}
\setcounter{enumiv}{\value{firstbib}}
\bibitem{qapone}
Daniel Funke, Sebastian Lamm, Peter Sanders, Christian Schulz, Darren Strash and Moritz von Looz.
\newblock {Communication-free Massively Distributed Graph Generation}.
\newblock In {\em 32nd IEEE International Parallel and Distributed Processing Symposium (IPDPS)}, pages 336--347, 2018. \textbf{Best Paper Award}. 

\bibitem{qapone}
Sebastian Schlag, Christian Schulz, Daniel Seemaier and  Darren Strash.
\newblock {Scalable Edge Partitioning}.
\newblock In {\em Proceedings of the 21th Workshop on Algorithm Engineering and Experimentation (ALENEX)}, pages 211--2225, SIAM, 2019.

\bibitem{abagenjv}
Peter Sanders and Christian Schulz.
\newblock {Scalable Generation of Scale-free Graphs}.
\newblock {\em Information Processing Letters}. Volume 116, Article No. 7, pages 489--491, 2016.

\bibitem{qapone}
Daniel Funke, Sebastian Lamm, Ulrich Meyer, Peter Sanders, Christian Schulz, Darren Strash and Moritz von Looz.
\newblock {Communication-free Massively Distributed Graph Generation}.
\newblock Invited to special issue of {\em Journal of Parallel and Distributed Computing for IPDPS'18}, to appear,~2019.

\setcounter{firstbib}{\value{enumiv}}
\end{thenewbibliography}
\section*{Memetic Algorithms}       
\begin{thenewbibliography}{100}
\setcounter{enumiv}{\value{firstbib}}
\bibitem{a}
Sebastian Lamm, Peter Sanders and Christian Schulz.
\newblock {Graph Partitioning for Independent Sets}.
\newblock In {\em Proceedings of the 14th Symposium on Experimental Algorithms (SEA)}, volume 8504 of Lecture Notes in Computer Science, pages 68--81. Springer, 2015.

\bibitem{a}
Nitin Ahuja, Matthias Bender, Peter Sanders, Christian Schulz and Andreas Wagner.
\newblock {Incorporating Road Networks into Territory Design}.
\newblock In {\em Proceedings of the 23rd International Conference on
Advances in Geographic Information Systems (GIS)}. ACM Press,~2015.
\bibitem{a}
Peter Sanders, Christian Schulz, Darren Strash and Robert Williger.
\newblock {Distributed Evolutionary $k$-way Node Separators}.
\newblock In {\em Proceedings of the Genetic and Evolutionary Computation Conference (GECCO)}, pages 345--252, 2017. 

\bibitem{qapone}
Robin Andre, Sebastian Schlag and Christian Schulz.
\newblock {Memetic Multilevel Hypergraph Partitioning}.
\newblock In {\em Proceedings of the Genetic and Evolutionary Computation Conference (GECCO)}, pages 347--354, ACM, 2018. 

\bibitem{qapone}
Orlando Moreira, Merten Popp and Christian Schulz.
\newblock {Evolutionary Multi-level Acyclic Graph Partitioning}.
\newblock In {\em Proceedings of the Genetic and Evolutionary Computation Conference (GECCO)}, pages 331--339, ACM, 2018.

\bibitem{qapone}
Sonja Biedermann, Monika Henzinger, Christian Schulz and Bernhard Schuster.
\newblock {Memetic Graph Clustering}.
\newblock In {\em Proceedings of the 17th Symposium on Experimental Algorithms (SEA)}, volume 103 of LIPIcs, pages 3:1--3:15, 2018. 

\setcounter{firstbib}{\value{enumiv}}
\end{thenewbibliography}

\newpage
\cleardoublepage
\chapter{Introduction}

Processing large complex networks recently attracted considerable interest. 
Complex graphs are useful in a wide range of applications from technological networks to biological systems like the human brain.
Sometimes these networks are composed of billions of entities that give rise to emerging properties and structures.
Analyzing these structures aids us in gaining new insights about our surroundings.
As huge networks become abundant, there is a need for scalable algorithms to perform analysis.
A prominent example is the PageRank algorithm, which is one of the measures
used by web search engines  such as Google 
to rank web pages displayed to the user. 
In order to find these patterns, massive amounts of data have to be acquired and processed.
Designing and evaluating \emph{scalable} graph algorithms to handle these datasets is a crucial task on the road to understanding the underlying systems.

\begin{figure}[t]
\includegraphics[width=\textwidth]{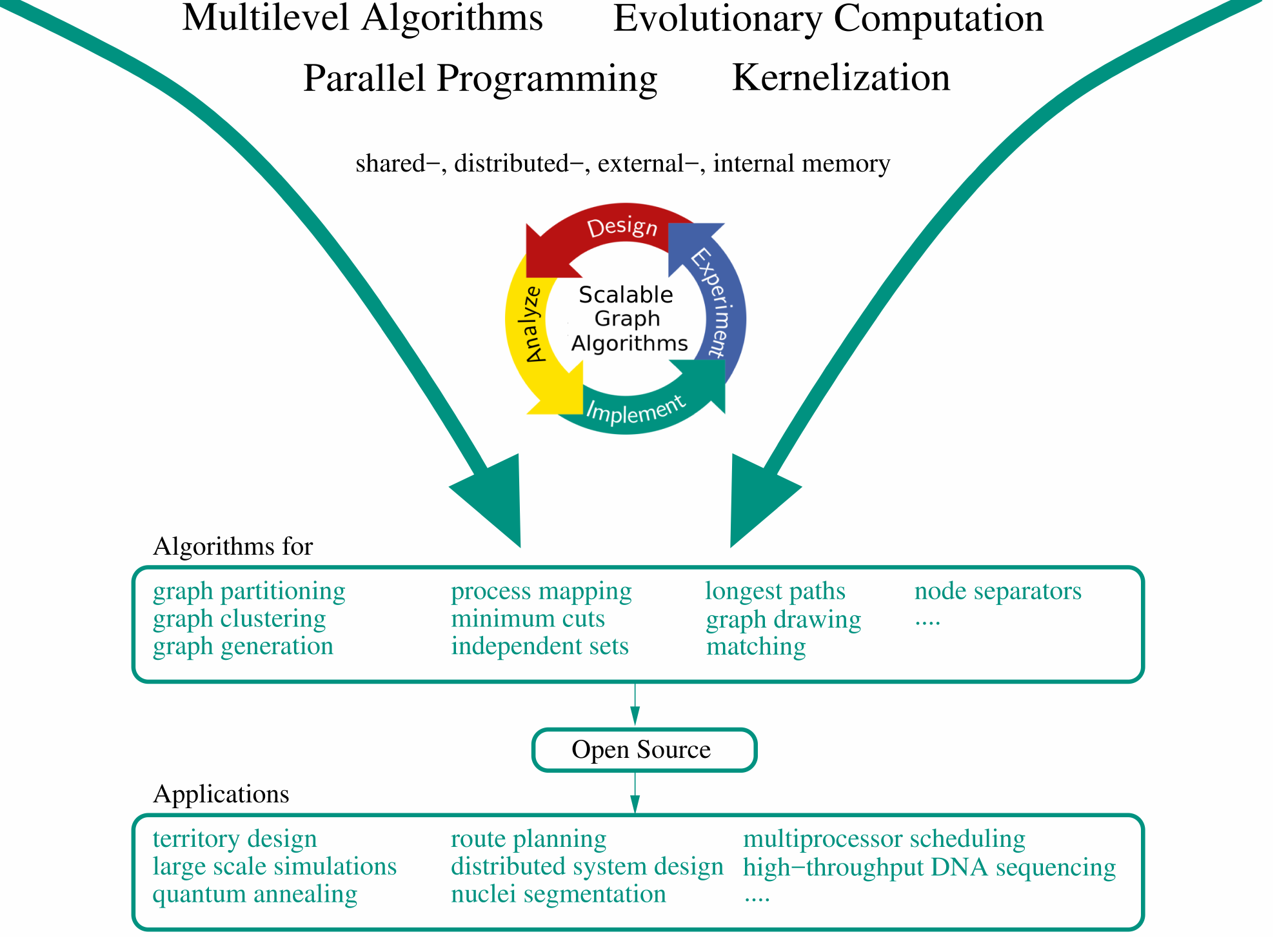}
\caption{Overview of the pillars and methodology used within this thesis.}
\label{fig:overviewfigure}
\end{figure}
This thesis tackles a broad spectrum of scalable graph algorithms. In general, this research is based on four pillars: \emph{multilevel algorithms}, \emph{practical kernelization}, \emph{parallelization} and \emph{memetic algorithms} that are highly interconnected.
Figure~\ref{fig:overviewfigure} gives an overview.
First, we look at multilevel algorithms for different graph problems.
All of these algorithms typically create a sequence of smaller graphs, compute a solution on the smallest graph and lastly transfer the solution through the computed hierarchy improving it on every level by making small adjustments. 
In particular, we design algorithms to partition complex networks, and then continue to look at structurally different problems, i.e.~multilevel algorithms for DAG partitioning, the node separator problem, the hypergraph partitioning problem, process mapping and graph~drawing.

Next, we investigate \emph{practical kernelization}. 
Many NP-hard graph problems have been shown to be fixed-parameter tractable (FPT): large inputs can be solved efficiently and provably optimally, as long as some problem parameter is small. 
Over the last two decades, significant advances have been made in the design and analysis of FPT algorithms for a wide variety of graph problems. 
This has resulted in a rich algorithmic toolbox that are by now well-established and are described in several textbooks and surveys. 
However, these theoretical algorithmic ideas have received very little attention from the practical perspective. 
Few of the new techniques are implemented and tested on real datasets, and their practical potential is far from understood. 
By applying techniques from data reduction algorithms in nontrivial ways, we obtain algorithms that perform surprisingly well on real-world instances for the independent set and the minimum cut problem. 
In general, we follow the scheme of iteratively applying exact and inexact reduction/kernelization rules to compute a much smaller problem kernel on which the problem under consideration is solved.

A lot of the graph algorithms proposed in this thesis fall into the \emph{parallelization} category. 
For example, the multilevel algorithms for graph drawing, graph partitioning, data reduction algorithms for the independent set problem or the minimum cut problem and some of the memetic algorithms that we propose are either shared-memory or distributed-memory parallel. On the other hand, graph partitioning and process mapping are a prerequisite to obtain scalable parallel graph algorithms.
We extend the toolbox of parallel graph algorithms with distributed-memory parallel algorithms that do not directly fit into the other research pillars. This includes parallel graph generation algorithms which are necessary to design and evaluate scalable graph algorithms on massively parallel machines  as well as distributed edge partitioning algorithms. 

Lastly, we develop several \emph{memetic algorithms}, i.e.~for the node separator problem, the hypergraph partitioning problem, the DAG partitioning problem, the territory design problem, as well as the graph clustering problem. These methods make heavy use of the already developed multilevel algorithms. 
However, we also develop a memetic algorithms, \ie for the independent set problem, which are not per se based on the multilevel paradigm but make use of the proposed multilevel graph partitioning techniques.

Note the pillars overall are \emph{highly interconnected} and the assignment of problems to pillars is not one-to-one, i.e.~some of the algorithms can be assigned~to~multiple~pillars.
For example, the multilevel algorithm algorithms iteratively reduce the problem size similar to kernelization algorithms with the difference that there is no guarantee to obtain a optimum solution.
On the other hand, applying exact data reduction rules to obtain an irreducible problem kernel and then solving the problem on the kernel, can also be viewed as a multilevel algorithm (without local search).
Generally, as explained above, we use parallelization everywhere possible. Lastly, the memetic algorithms are based on the previously engineered multilevel algorithms and also partially use data reduction rules to speed up memetic algorithms in practice.

The \emph{main methodology} used in this thesis is the \emph{algorithm
  engineering} approach to algorithmics.
Traditionally, algorithms are designed and analyzed using simple models of problems and machines.
This often yields important performance guarantees for all possible inputs in terms of running time or solution quality that an optimization algorithm can achieve.
Hence, algorithms are usable in  many
applications with predictable performance for previously unknown inputs. 
In algorithm theory, however, taking up and implementing an algorithm
is part of the application development which has to be done by the person that wants to use the algorithm. 
Unfortunately, this mode of transferring results is a slow process and sometimes the theoretically best algorithms perform poor in practice due to large constant factors that are typically hidden the analysis. There is also a growing gap between theory
and practice: Realistic hardware with its parallelism, memory hierarchies etc.~is
diverging from traditional machine models. 
 In contrast to algorithm theory, algorithm engineering uses an innovation cycle where
algorithm design based on realistic models, theoretical analysis, efficient
implementation, and careful experimental evaluation using real-world inputs.
This closes gaps between theory and practice and leads to improved application codes
and reusable software libraries (see {\small\url{www.algorithm-engineering.de}}). This yields results that practitioners can rely on for their specific application. 

Within this thesis we engineer algorithms that
improve known methods especially for large instances.
We released the techniques and algorithms that have been developed in work that forms the base of this thesis as easy-to-use open source software.  
This includes a
widely used library of algorithms for graph
partitioning~\cite{kahipWebsite}, graph drawing~\cite{kadrawWebsite},
independent sets~\cite{kamisWebsite}, hypergraph partitioning~\cite{kahyparWebsite}, graph clustering~\cite{vieclusWebsite}, graph generation~\cite{kagenWebsite}, minimum cuts~\cite{viecutWebsite} and process mapping~\cite{viemaWebsite}. 
Hence, the concrete outcome of this habilitation thesis are well-engineered, usable
systems that are more robust, more flexible, produce better
solutions, and scale to massively parallel machines and instances much
larger than previously possible.  En passant, we arrived at
more scalable algorithms for a wide range of graph problems. 

To give a couple of \emph{highlights}: our parallel graph partitioning algorithms for complex networks can
compute a partition of graph with billions of edges in only a few seconds while producing much better solutions than competing algorithms. The independent set algorithms developed in this thesis~\cite{evoIS,kerMIS,MISonthefly,misatscalejv,parallelmis,exactWMIS} find large independent sets much faster than existing local search algorithms, are competitive with state-of-the-art exact algorithms for smaller graphs, and allow us to compute large independent sets on huge sparse graphs, with billions of edges.
In addition, our new algorithm computes an optimal independent set on all instances that the exact algorithm can solve.  
Our graph clustering algorithm~\cite{clustering} is able to reproduce or improve previous all entries of the 10th DIMACS implementation challenge under consideration as well as results recently reported in the literature in a short amount of time.  Moreover, while the previous best result for different instances has been computed by a variety of solvers, our algorithm can now be used as a single tool to compute the result.
The minimum cut algorithms~\cite{mincutjv,smexact} presented in this thesis are the fastest currently available for the problem beating the previous fastest algorithm by an order of magnitude.
As a last example: our parallel graph generation algorithms~\cite{bagenjv,graphgennew,jvgraphgennew} \emph{require no communication at all} and hence have perfect load balance on uniform nodes of a supercomputer. Hence, our algorithms achieve almost perfect scalability.
We generated a Petaedge graph in less than an hour on 16\,384 cores of the SuperMUC computer.
This graph is 20\,000 times larger than the largest Barabasi-Albert graph we have seen reported.

\chapter{Multilevel Algorithms}
\label{sec:multilevel}
In the first pillar of research that we outline, we look at multilevel algorithms for different graph problems.
All of these algorithms typically create a sequence of smaller graphs, compute a solution on the smallest graph and lastly transfer the solution through the computed hierarchy improving it on every level by making small adjustments. 
The intuition of the multilevel scheme is that a good solution at one level of the hierarchy will also be a good solution on the next finer level. 
Then local search has a more global view on the problem on the coarse levels and a very fine-grained view on the fine levels of the multilevel hierarchy.
Hence, 
depending on the definition of the neighborhood, local search algorithms are able to explore local solution spaces very effectively~in~this~setting. 
The section is structured as follows: we start with multilevel algorithms designed to partition complex networks, and then continue to look at structurally different problems, i.e.~multilevel algorithms for DAG partitioning and for the node separator as well as the hypergraph partitioning problem. We finish this section by outlining multilevel algorithms for process mapping and graph~drawing.

\section{Partitioning Highly Irregular Networks}
\label{multilevelpart}
\mycoauthor{This section is based on \cite{clustersnw, externalcomplexnetworks,parallelsnw,sharedmempart, clustersnwjv, parallelsnwjv} which are joint publications with Yaroslav Akhremtsev, Henning Meyerhenke and Peter Sanders.}
\noindent Graph partitioning (GP) is important for processing very large graphs, \eg networks stemming from finite element methods, route
planning, social networks or web graphs.
Often the node set of such graphs needs to be partitioned such that there are few edges between the blocks (node subsets, parts).  In
particular, when you process a graph in parallel on $k$ PEs (processing
elements), you often want to partition the graph into $k$ blocks of (about) equal
size. Then each PE owns a roughly equally sized part of the graph.
In this paper we focus on a commonly used version of the problem that constrains the
maximum block size to $(1+\epsilon)$ times the average block size and tries to
minimize the total cut size, i.e., the number of edges that run between blocks.
Such edges are supposed to model the communication at block boundaries between the corresponding PEs.
It is well-known that there are more realistic (but more complicated) objective
functions involving also the block that is worst and the number of its
neighboring nodes~\cite{HendricksonK00}, but the cut size has been the predominant optimization criterion.
The GP decision problem is NP-complete and there is no approximation algorithm with a constant factor for general graphs~\cite{BuiJ92}. Thus heuristic algorithms are used in practice. 

Complex networks, such as social networks or web graphs, often feature a set of clusters organized in a hierarchical fashion~\cite{lancichinetti2009detecting}.
Such graphs have become a focus of
investigation~\cite{costa2011analyzing} and target for graph partitioning. While partitioning meshes is a 
mature field, the structure of complex networks poses new challenges to graph
partitioning methods. Complex networks are often \emph{scale-free} (many low-degree
nodes, few high-degree nodes) and have the {small-world property}.
Small world means that the network has a small
diameter, so that the whole graph is discovered within a few hops from any source node.
These two properties distinguish complex networks from traditional meshes and make finding
small cuts difficult with established tools.
Yet, to cope with massive network data sets in reasonable time, there is a need 
for parallel algorithms. Their efficient execution requires the partitioning of such networks
onto different processing elements (PEs) with high quality.
\begin{figure}[t!]
\begin{center}
\subfigure[]{
\includegraphics[width=0.61\textwidth]{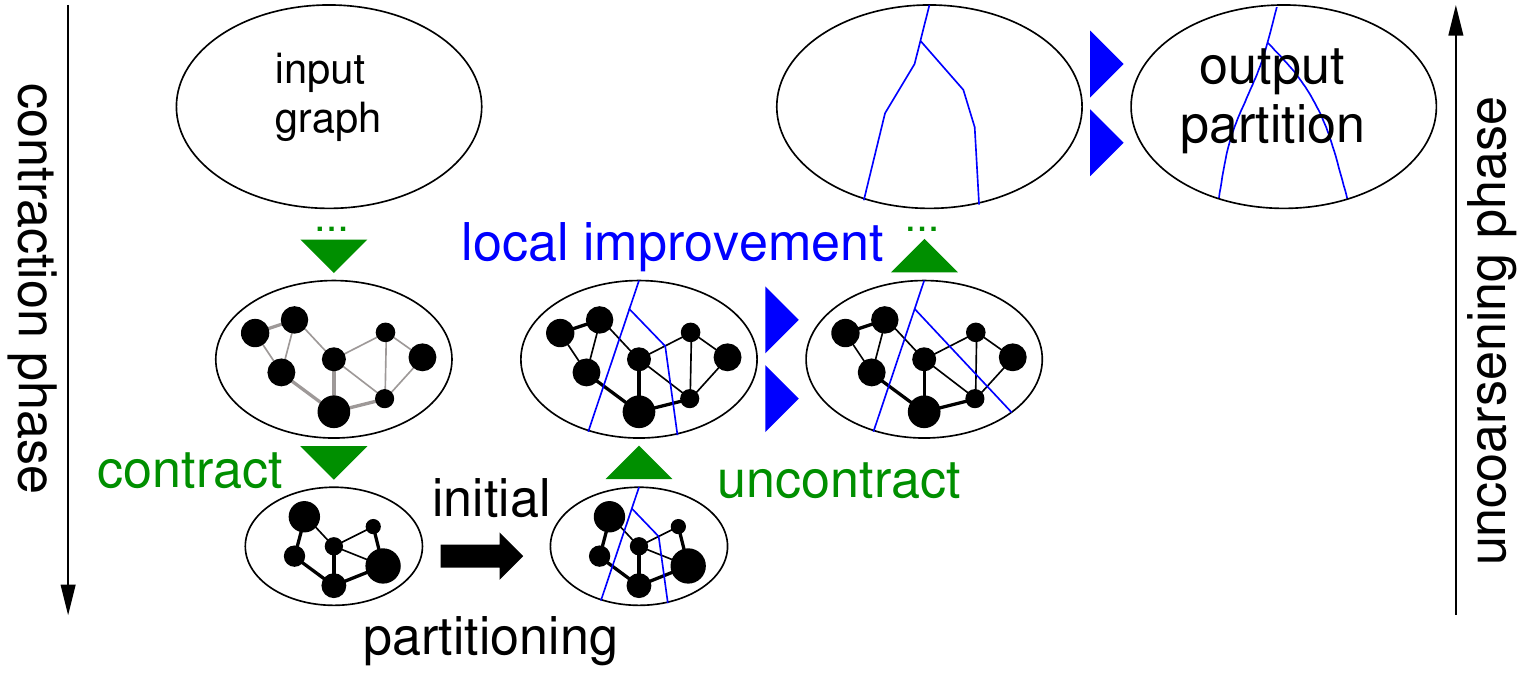}
}
\hfill
\subfigure[]{
\includegraphics[width=0.33\textwidth]{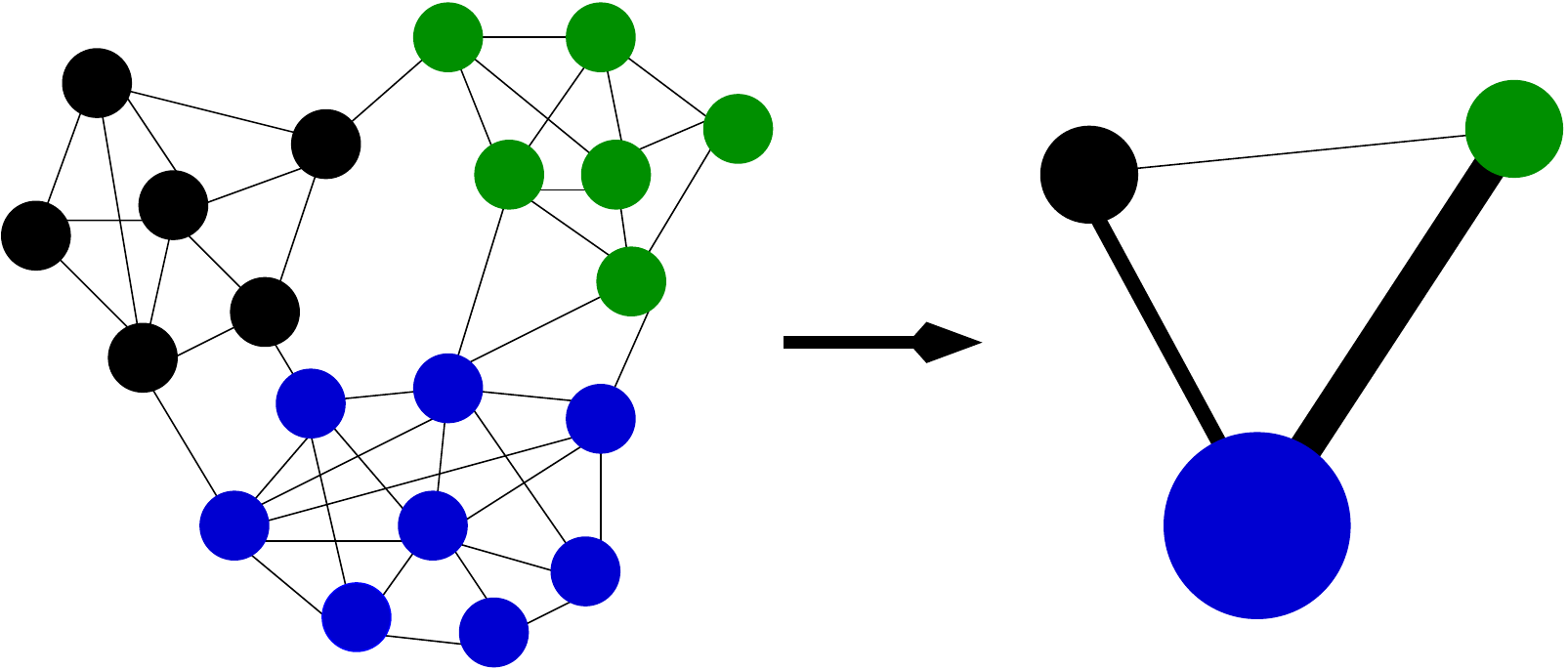}
}
\caption{(a) Sketch of multilevel graph partitioning, the predominant GP heuristic framework in practice.
(b) Contraction of a clustering. The clustering of the graph is indicated by the colors. Each cluster of the 
graph left of the arrow corresponds to a node (= supervertex) in the graph right of the arrow.}
\label{fig:mgp}
\label{fig:clustercontraction}
\end{center}
\end{figure}

In \cite{clustersnw, clustersnwjv} we present a new multilevel algorithm designed for partitioning complex networks. Its rationale 
is to use aggressive cluster-based coarsening and simple, yet effective local search.  Our new \underline{S}ize-\underline{C}onstrained \underline{La}bel \underline{P}ropagation
(SCLaP) algorithm serves both purposes. During coarsening, we compute a graph clustering with 
size-constrained clusters and contract each cluster to a supervertex, also see Figure~\ref{fig:clustercontraction}b.
This contraction yields a new 
level in the multilevel hierarchy. During uncoarsening the same algorithm can be used as a fast local 
search algorithm.

To perform graph clustering with size constraints on the clusters, we propose a size-constraint label propagation algorithm. 
The \emph{label propagation algorithm} (LPA), originally proposed by Raghavan \etal \cite{labelpropagationclustering},
fulfills these requirements, with the exception that it does not constrain cluster sizes.
It is a fast, near-linear time algorithm that locally optimizes the number of edges cut. We outline the algorithm briefly.  
Initially, each node is in its own cluster, \ie the initial cluster ID of a node is set to its node ID.
The algorithm then works in rounds. In each round, the nodes of the graph are traversed in random order. 
When a node $v$ is visited, it is \emph{moved} to the block that has the strongest connection to $v$, \ie it is moved 
to the cluster $V_i$ that maximizes $\omega(\{(v, u) \mid u \in N(v) \cap V_i \})$. Ties are broken randomly. 
The process is repeated for at most $\ell$ iterations where $\ell$ is a tuning parameter. 

Furthermore, we augment the basic SCLaP algorithm by several algorithmic components that are supposed
to improve speed and/or quality of the partitioning process.
These extensions include different orders of node traversals, combining several clusterings into one,
iterations of the multilevel approach, larger imbalance on coarser levels, and the omission of nodes in the
clustering process that are certain not to change their cluster.
Our experiments on various complex networks indicate that the presented algorithm, more precisely its 
different configurations integrated into the partitioning tool KaHIP, is clearly able to improve on the state of the art.
Excellent partitioning quality is obtained in a short amount of time.
For example, a web graph with 3.3 billion edges can be partitioned with sequential code in about ten minutes 
while cutting less than half of the edges than the partitions computed by the established competitor kMetis.

Subsequently, we adapted the algorithm to the distributed-memory model~\cite{parallelsnw,parallelsnwjv}.
Our main contributions here are a scalable parallelization of the size-constrained label propagation algorithm and an integration into a multilevel framework that enables us to partition large complex networks on a parallel machine. 
The parallel size-constrained label propagation algorithm is used to compute a graph clustering. A clustering of this kind
is recursively contracted and recomputed on the coarser graph until the coarsest graph is small enough. 
The coarsest graph is then partitioned by the coarse-grained distributed evolutionary algorithm KaFFPaE~\cite{kaffpaE}.
During uncoarsening the size-constrained label propagation algorithm is used as a simple, yet effective, parallel local search algorithm. 

Our parallel label propagation algorithm uses a parallel graph data structure such that each processors holds a subgraph of the input.
To parallelize the label propagation algorithm, each PE performs the algorithm on its part of the graph. 
Recall, when we visit a node $v$, it is moved to the block that has the strongest eligible connection.
Note that the cluster IDs of a node can be arbitrarily distributed in the range $0\, .. \, n-1$ so that we use a hash map to identify the cluster with the strongest connection.
Our algorithm uses the following scheme to \emph{overlap} communication and computation. 
The scheme is organized in phases. 
We call a node \emph{interface node} if it is adjacent to at least one ghost node (\ie a node that is in the subgraph of another processor). The PE associated with the ghost node is called adjacent PE.
Each PE stores a separate send buffer for all adjacent PEs.
During each phase, we store the block ID of interface nodes that have changed into the send buffer of each adjacent PE of this node. 
Communication is then implemented \emph{asynchronously} using non-blocking  operations, \ie while messages are routed through the network, we compute new labels of the nodes of the next phase. 
In phase $\kappa$, we send the current updates to the adjacent PEs and receive the updates of the adjacent PEs from round $\kappa-1$, for $\kappa>1$. 
Note that in case the label propagation algorithm has converged, \ie no node changes its block any more, the communication volume is really small.

The presented scheme speeds up the running time needed for partitioning and improves solution quality on graphs that have a very irregular and often also hierarchically clustered structure such as social networks or web graphs.
On these graphs the \emph{strengths} of our new algorithm unfold in particular: average solution quality \emph{and} running time is much better than what is observed by using ParMetis~\cite{karypis1996parallel}.
A variant of our algorithm is able to compute a partition of a web graph with billions of edges in only a few seconds while producing much better solutions.

It turns out that the main idea is also feasible for the (semi)-external memory model~\cite{externalcomplexnetworks} and shared-memory parallel model~\cite{sharedmempart}.
The algorithm that runs in the semi-external model~\cite{externalcomplexnetworks} is able to partition and cluster huge complex networks with billions of edges on cheap commodity machines.
Moreover, we present the \emph{first} fully external graph clustering/partitioning algorithm that is able to deal with a size-constraint on the blocks.
Experiments demonstrate that the semi-external graph partitioning algorithm is scalable and can compute high quality partitions in time that is comparable to the running time of an efficient internal memory implementation. 
Lastly, we engineered a high-quality shared-memory parallel algorithm which has a stronger focus high solution quality~\cite{sharedmempart}. 
We present an approach to multilevel shared-memory parallel graph partitioning that guarantees balanced solutions, shows high speed-ups for a variety of large graphs and yields very good quality independently of the number of cores used.
For example, on 31 cores, our algorithm partitions our largest test instance 
into 16 blocks cutting \emph{less than half} edges than our main
competitor when both algorithms are given the same amount of time.
Important ingredients include parallel label propagation,
parallel initial partitioning, a simple yet effective approach to parallel localized local search, and cache-aware hash tables.

\section{DAG Partitioning}
\mycoauthor{This section is based on \cite{dagone,evoDAG} which are joint publications with Orlando~Moreira and Merten Popp.}
\noindent The context of this research is the development of computer vision and imaging
applications at Intel Corporation. These applications have high demands for
computational power but often need to run on embedded devices with severely
limited compute resources and a tight thermal budget. Our target platform is a
heterogeneous multiprocessor for advanced imaging and computer vision that is
currently used in Intel processors. It is designed for low power and has small
local program and data memories. To cope with memory constraints, the
application developer currently has to \emph{manually} break the application, which is given as
a directed data flow graph, into smaller blocks that are executed one after another. The
quality of this partitioning has a strong impact on communication volume and
performance. However, for large graphs this is a non-trivial task that requires
detailed knowledge of the hardware.

There are many existing heuristics for partitioning graphs into blocks of nodes
of roughly equal size. However, the platform at Intel has the requirement that there
must not be a cycle in the dependencies between the blocks because they have to
be executed one after another.
The task can be identified as a graph partitioning problem. The partitions that we are looking for have to satisfy two constraints: a
balancing constraint and an acyclicity constraint. The \emph{balancing
constraint} demands that
$\forall i\in \{1..k\}\gilt c(V_i) \leq L_{\max} := (1+\epsilon)\lceil\frac{c(V)}{k}\rceil$
for some imbalance parameter $\epsilon \geq 0$.
The \emph{acyclicity constraint} mandates that
the quotient graph is acyclic.
The objective is to minimize the total \emph{cut} $\sum_{i,j}w(E_{ij})$ where
$E_{ij}\Is\setGilt{(u,v)\in E}{u\in V_i,v\in V_j}$.
The \emph{directed graph partitioning problem with acyclic quotient graph (DGPAQ)}
is then defined as finding a partition $\varPi:=\left\{V_{1,}\ldots,V_{k}\right\}$
that satisfies both constraints while minimizing the objective function.

In \cite{dagone} we present a proof that the problem is NP-complete and hard
to approximate, as well as the implementation and evaluation of heuristics that address
this problem. Roughly speaking, we generate initial feasible solutions by exploiting topological orderings and later on improve solutions by using local search algorithms that do not create cycles in the quotient graph.
Our experiments show that our heuristics achieve  a close approximation of the optimal solution found by an exhaustive search
    for small problem instances and much better scalability for larger
    instances.
    Then we extend our idea to a multilevel scheme~\cite{evoDAG}. In principle, we start with a feasible solution and modify the coarsening process to only contract edges that run entirely within a block. This ensures that we can use the input solution as feasible initial solution on the coarsest level and afterwards apply local search on every level of the graph hierarchy.  Based on the multilevel algorithms defined here, we contributed an memetic algorithm for the problem (see Section~\ref{evolutionary}).
    The resulting partitions have shown to have a positive impact on the schedule of a real imaging
    application. 
\vfill
\pagebreak
\section{Node Separators}
\mycoauthor{This section is based on \cite{NS,evoNS} which are joint publications with Peter Sanders, Darren Strash and Robert Williger.}
\noindent A node separator of a graph is a subset $S$ of the nodes such that removing $S$ and its incident edges divides the graph into two disconnected components of about equal size.
There are many algorithms that rely on small node separators. 
For example, small balanced separators are a popular tool in divide-and-conquer strategies~\cite{lipton1980applications,leiserson1980area,BHATT1984300}, are useful to speed up the computations of shortest paths~\cite{schulz2002using,delling2009high,dibbelt2014customizable} or are necessary in scientific computing to compute fill reducing orderings with nested dissection algorithms~\cite{george1973nested}.

Finding a balanced node separator on general graphs is NP-hard even if the maximum node degree is three~\cite{bui1992finding,garey2002computers}. 
Hence, one relies on heuristic and approximation algorithms to find small node separators in general graphs.
Indeed, one of the most commonly used method to tackle the node separator problem on large graphs in practice is the multilevel approach.

In \cite{NS}, we introduce novel multilevel algorithms to find small node separators in large graphs. 
With focus on solution quality, we introduce novel flow-based local search algorithms which are integrated in a multilevel framework. 
In addition, we transfer techniques successfully used in the graph partitioning field. 
This includes the usage of edge ratings tailored to our problem to guide the graph coarsening algorithm as well as  highly localized local search and iterated multilevel cycles to improve solution quality even further. 
The two most important ingredients here are highly localized local search and flow-based local search.

The novel localized algorithm for the node separator problem starts local search only from a couple of selected separator nodes. 
Our localized local search procedure is based on the FM scheme. 
Before we explain our approach to localization, we present a commonly used FM-variant for completeness. 
For each of the two blocks $V_1$, $V_2$ under consideration, a priority queue of separator nodes eligible to move is kept. 
The priority is based on the \emph{gain} concept, \ie the decrease in the objective function value when the separator node is moved into that block. 
More precisely, if a node $v \in S$ would be moved to $V_1$, then the neighbors of $v$ that are in $V_2$ have to be moved into the separator. 
Hence, in this case the gain of the node is the weight of $v$ minus the weight of the nodes that have to be added to the separator. 
The gain value in the other case (moving $v$ into to $V_2$) is similar. 
After the algorithm computed both gain values it chooses the largest gain value such that moving the node does not violate the balance constraint and performs the movement.
Each node is moved at most once out of the separator within a single local search. 
The queues are initialized randomly with the separator nodes. 
After a node is moved, newly added separator nodes become \emph{eligible} for movement (and hence are added to the priority queues). The moved node itself is not eligible for movement anymore and is removed from the priority queue. 
Note that the movement can change the gain of current separator nodes. Hence, gains of adjacent nodes are updated.

Our approach to localization works as follows. 
Previous local search methods were initialized with \emph{all} separator nodes, \ie all separator nodes are \emph{eligible} for movement at the beginning.
In contrast, our method is repeatedly initialized only with a \emph{subset} of the separator nodes (the precise amount of nodes in the subset is a tuning parameter).  
Intuitively, this introduces a larger amount of diversification and boosts the algorithms ability to climb out of local minima.

We now give intuition why localization of local search boosts the algorithms ability to climb out of local minima.
Consider a situation in which a node separator is locally optimal in the sense that at least two node movements are necessary until moving a node out of the separator with positive gain is possible. Recall that classical local search is initialized with all separator nodes (in this case all of them have negative gain values). 
It then starts to move nodes with negative gain at multiple places of the graph. 
When it finally moves nodes with positive gain the separator is already much worse than the input node separator.
Hence, the movement of these positive gain nodes does not yield an improvement with respect to the given input partition. 
On the other hand, a localized local search that starts close to the nodes with positive gain, can find the positive gain nodes by moving only a small number of nodes with negative gain. 
Since it did not move as many negative gain nodes as the classical local search, it may still finds an improvement with respect to the input.

\begin{figure}[t]
\begin{center}
\includegraphics[width=250pt]{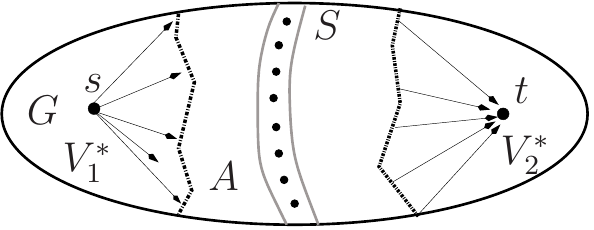}
\end{center}
\caption{ The construction of an $A$ induced flow problem $\mathcal{F}$ is shown. 
        Two BFS are started to define the area $A$ -- one into the block on the left hand side and one into the block on the right hand side.
        A solution of the flow problem yields the smallest node separator that can be found within the area.  The area $A$ is chosen so that each node separator being found in it yields a feasible separator for the original problem.}
        \label{fig:flowconstruction}
\end{figure}

The flow-based improvement methods work as follows; We define a node-capacitated flow problem $\mathcal{F}=(V_\mathcal{F}, E_\mathcal{F})$ that we solve to improve a given node separator as follows.
First we introduce a few notations. 
Given a set of nodes $A \subset V$, we define its \emph{border}~$\partial A := \{ u \in A \mid \exists (u,v) \in E : v \not\in A\}$.
The set $\partial_1 A := \partial A \cap V_1$ is called \emph{left border} of $A$ and the set $\partial_2 A := \partial A \cap V_2$ is called \emph{right border} of $A$. 
An \emph{$A$ induced flow problem}~$\mathcal{F}$ is the node induced subgraph $G[A]$ with $\infty$ as edge-capacities and the node weights of the graph as node-capacities. Additionally there are two nodes $s,t$ that are connected to the border of $A$. 
More precisely, $s$ is connected to all left border nodes~$\partial_1 A$ and all right border nodes $\partial_2 A$ are connected to $t$.  
These new edges get capacity $\infty$. 
Note that the additional edges are directed.
$\mathcal{F}$ has the \emph{balance property} if each ($s$,$t$)-flow induces a balanced node separator in $G$, \ie the blocks $V_i$ fulfill the balancing constraint.
The basic idea is to construct a flow problem $\mathcal{F}$ having the balance property. Finding the vertices for the flow problem can be done by running two breadth first search starting from the separator (one for each block, where the breadth first search only expands into block $V_i$). An example is shown in Figure~\ref{fig:flowconstruction}.

Experiments indicate that flow-based local search algorithms on its own in a multilevel framework are already \emph{highly} competitive in terms of separator quality. 
Adding additional local search algorithms further improves solution quality.
Our strongest configuration almost always outperforms competing systems while on average computing 10\% and 62\% smaller separators than Metis and Scotch (the main competitors), respectively.
Later on we developed algorithms for the $k$-way node separator problem~\cite{evoNS}.
Parts of the algorithms have already been integrated in the KaHIP framework.
Based on the multilevel algorithms defined here, we also contributed an memetic algorithm for the problem (see Section~\ref{evolutionary}).

\section{Hypergraph Partitioning}

\mycoauthor{This section is based on \cite{HGBisecNLevelTR,evoHG} which are joint publications with Robin Andre, Vitali Henne, Tobias Heuer, Henning Meyerhenke, Peter Sanders and Sebastian~Schlag.}

Hypergraphs are a generalization of graphs, where each (hyper)edge can connect more than two vertices.
The $k$-way hypergraph partitioning problem is the generalization of the well-known graph partitioning problem:
Partition the vertex set into $k$ disjoint blocks of bounded size
(at most $1+\varepsilon$ times the average block size), while minimizing the total cut size, i.e., the sum of the weights of those hyperedges that connect multiple blocks.
However, allowing hyperedges of arbitrary size makes the
partitioning problem more difficult in practice~\cite{Application:DistributedDB,Heintz:2014}.

Hypergraph partitioning (HGP) has a wide range of applications. Two prominent areas are 
VLSI design and scientific computing (\eg accelerating sparse matrix-vector multiplications)~\cite{Papa2007}.
While the former is an example of a field where small optimizations can lead to significant savings, the latter
is an example where hypergraph-based modeling is more flexible than graph-based approaches~\cite{Heintz:2014,PaToH,HendricksonK00,10.1371/journal.pcbi.1000385}. HGP also finds application as a preprocessing step in SAT solving, where it is used to identify groups of connected variables~\cite{DBLP:journals/jucs/AloulMS04}.

State-of-the-art hypergraph partitioners use matching- or clustering-based algorithms to find 
groups of highly-connected vertices that can be contracted together to create the next level of the 
coarsening hierarchy~\cite{PaToH,hMetisRB,trifunovic2006parallel}. The rate at which successively coarser hypergraphs are reduced determines 
the number of levels in the multilevel hierarchy. As already noted in \cite{MLPart}, a larger number of levels potentially improves the solution quality, 
because local search algorithms are used more often. However, it also leads to larger running times and increased memory usage.

We develop a multilevel algorithm for hypergraph partitioning \cite{HGBisecNLevelTR} that contracts the vertices one at a time. Using several caching and lazy-evaluation techniques during coarsening and refinement, we reduce the running time by up to two-orders of magnitude compared to a naive $n$-level algorithm that would be adequate for ordinary graph partitioning. The overall performance is even better than the widely used hMetis hypergraph partitioner that uses a classical multilevel algorithm with few levels.

Roughly speaking our overall system to achieve very high quality works as follows: A carefully chosen rating function evaluates how attractive it is to 
contract two vertices. We present an effective strategy to limit the cost for reevaluating the rating function and always contract the pair of vertices having the largest rating. When coarsening is stopped, we run our initial partitioner. 
The initial partitioner is based on a large portfolio of simple algorithms, each with some randomization aspect (fully random, BFS, label propagation, and nine variants of greedy hypergraph growing). 
Since these partitioners are very fast and only applied to a small core problem, we can afford to make a large number of attempts taking the best partition as the basis for further processing. 
Lastly, local improvement steps are expensive since many steps are needed and a naive implementation of the established techniques needs work proportional to the \emph{squares} of the net sizes. We integrate several techniques for reducing this bad behavior for large nets and additionally develop a way to \emph{cache} gain values to further reduce search overhead.

We assembled a large benchmark set with 310 hypergraphs stemming from
application areas such VLSI, SAT solving, social networks, and scientific computing. Averaged over all instances, we achieve about 8\% smaller cuts than hMetis and are at the same time about twice as fast.
Compared to the (much faster) partitioner PaToH, we achieve cuts that are about 4\% smaller. Considerably larger improvements are observed for some instance classes like social networks and for bipartitioning.
The algorithm presented forms the basis of the hypergraph
partitioning framework \emph{KaHyPar} (\textbf{Ka}rlsruhe \textbf{Hy}pergraph \textbf{Par}titioning).
Based on this we contributed an memetic algorithm~\cite{evoHG} for the problem (see Section~\ref{evolutionary}).These algorithms have been released in the KaHyPar~\cite{kahyparWebsite}-- Karlsruhe Hypergraph Partitioning -- framework.

\vfill
\pagebreak
\section{Process Mapping}
\mycoauthor{This section is based on \cite{processmapping} which is joint work with Jesper Larsson Träff.}

\emph{Process mapping} is an important generalization of graph partitioning for massively parallel computing: How to embed a (very) large graph into the network of a supercomputer so as to minimize communication costs. Although this problem has been considered by multiple researchers already, there is still a long way to go to find really adequate solutions. 
Communication performance between processes in high-performance
systems depends on many factors. For example, communication is
typically faster if communicating processes are located on the same
processor node compared to the cases where processes reside on
different nodes.  This becomes even more pronounced for large
supercomputer systems where processors are hierarchically organized
into, \eg islands, racks, nodes, processors, cores with corresponding
communication links of similar quality.  Given the communication
pattern between processes and a hardware topology description that
reflects the quality of the communication links, one hence seeks to
find a good mapping of processes onto processors such that pairs of
processes exchanging large amounts of information are located closely.

We addressed the problem as quadratic assignment problem and present algorithms to construct initial mappings of processes to processors as well as fast local search algorithms to further improve the mappings~\cite{processmapping}. By exploiting assumptions that typically hold for applications and modern supercomputer systems such as sparse communication patterns and hierarchically organized communication systems, we arrive at significantly more powerful algorithms for these special QAPs. We outline algorithm to construct good mappings:

Intuitively, we want to identify subgraphs in the
communication graph of processes that have to communicate much with
each other and then place such processes closely, \ie on the same
node, same rack and so forth.  In the following, we assume a
homogeneous hierarchy of the supercomputer, but our algorithms can be extended to heterogeneous hierarchies in a straightforward way. Let $\mathcal{S}=a_1, a_2,
..., a_k$ be a sequence describing the hierarchy of the
supercomputer. The sequence should be interpreted as each processor having
$a_1$ cores, each node $a_2$ processors, each rack $a_3$ nodes, \ldots.
Our algorithm, called \emph{top down}, splits the communication graph recursively along the system hierarchy.

The \emph{top down approach} starts by computing a \emph{perfectly balanced} partition of $G_\mathcal{C}$ into $a_k$ blocks each having $n/a_k$ vertices (processes). Here, $G_\mathcal{C}$ is the graph describing the communication that takes place in the application where each vertex models a task that has to be placed on a single processor and edge weights model communication volume between the tasks. The partitioning task is done using the techniques provided by Sanders and Schulz~\cite{kabapeE} which provide high quality partitions and guarantee that each block of the output partition has the specified amount of vertices. In principle, the nodes of each block will be assigned completely to one of the $a_k$ system entities. 
Each of the system entities provides precisely $n/a_k$ PEs. 
We then proceed recursively and partition each subgraph induced by a block into $a_{k-1}$ blocks and so forth. 
The recursion stops as soon as the subgraphs have only $a_1$ vertices left. 
In the base case, we assign processes to permutation~ranks.

Experiments indicate that our algorithms find much better solutions and are much more scalable than the state-of-the-art. The algorithms have already been integrated in the KaHIP framework to map the blocks of a partition to processors as well as released separately in the VieM \cite{viemaWebsite} (Vienna Mapping and Sparse Quadratic Assignment) framework to make the mapping algorithm available in a more general context.

\section{Graph Drawing}
\mycoauthor{This section is based on \cite{drawing,jvdrawing} which are joint publications with Henning Meyerhenke and Martin Nöllenburg.}

Drawing large networks (or graphs, we use both terms interchangeably) with hundreds of thousands of vertices and
edges has a variety of relevant applications. One of them can be interactive visualization, which helps humans working on
graph data to gain insights about the properties of the data. If a very large high-end display is not available for such
purpose, a hierarchical approach allows the user to select an appropriate zoom level~\cite{abello-ask-06} in the spirit of Shneiderman's information seeking mantra ``Overview first, zoom and filter, then details-on-demand''~\cite{s-ehtdttiv-96}.
Moreover, drawings of large graphs can also be used as a preprocessing step in high-performance applications~\cite{Kirmani:2013:SPG:2503210.2503280}.

One very promising class of layout algorithms in this context is based on the
\emph{stress} of a graph, an error measure between true distances and desired target distances of vertex pairs. Such algorithms can, for instance, be used for drawing graphs with given
distances between vertex pairs, provided \textit{a priori} in a distance matrix~\cite{gkn-gdsm-05}.
More recently, Gansner \etal\cite{ghn-mmgl-13} proposed a similar model 
that includes besides the stress for some vertex pairs an additional entropy term to resolve the remaining degrees of freedom in the layout (hence its name \emph{maxent-stress}). 
While still using shortest path distances, this model often results in more satisfactory layouts for large networks.
The optimization problem can be cast as solving Laplacian 
linear systems successively. Since each right-hand side in this succession 
depends on the previous solution, many linear systems need
to be solved until convergence.

Gansner \etal\cite{ghn-mmgl-13} also suggested (but did not use) a simpler iterative refinement procedure for solving 
their optimization problem. This Jacobi method would be slow to converge if used unmodified. However, if designed and 
implemented appropriately, it has the potential for fast convergence even on large graphs. Moreover, as already observed
in~\cite{ghn-mmgl-13}, it has high potential for parallelism and should work well on dynamic graphs by profiting from  
previous solutions.

A successful (meta)heuristic for graph drawing (and other optimization problems on large graphs) is the multilevel approach. We also employ this approach for maxent-stress optimization for several other reasons:
(i) Some graphs (such as road networks) feature a hierarchical structure, which can be exploited to some extent
by a multilevel approach, (ii) the computed hierarchy may be useful later on for multiscale visualization, and
(iii) it resembles the multigrid method for solving linear equations.

We make the iterative local optimizer suggested by Gansner \etal\cite{ghn-mmgl-13}  
usable and fast in practice \cite{drawing,jvdrawing}. 
We briefly sketch our algorithmic 
approach: The method for creating the graph hierarchy is based on fast graph clustering with 
controllable cluster sizes (we use size-constraint label propagation). 
Each cluster computed on one hierarchy level is contracted into a new supervertex for the next level.
We denote vertices of contracted graphs in the hierarchy as \emph{nodes} to distinguish them from the original vertices of the input graph $G$.
After computing an initial layout on the coarsest hierarchy level, we improve the drawing 
on each finer level by using an iterative equation. 
One property of the local optimizer we exploit is its high degree of parallelism (which we use). 
Additionally, this refinement process exploits
the hierarchy and draws nodes that are densely connected with each other (\ie which are in the same cluster) close to each other. Further acceleration is obtained by approximating long-range forces. 
To this end, we use coarser representatives stored in the multilevel hierarchy.

\begin{figure*}[tb]
\centering
\includegraphics[width=4.25cm]{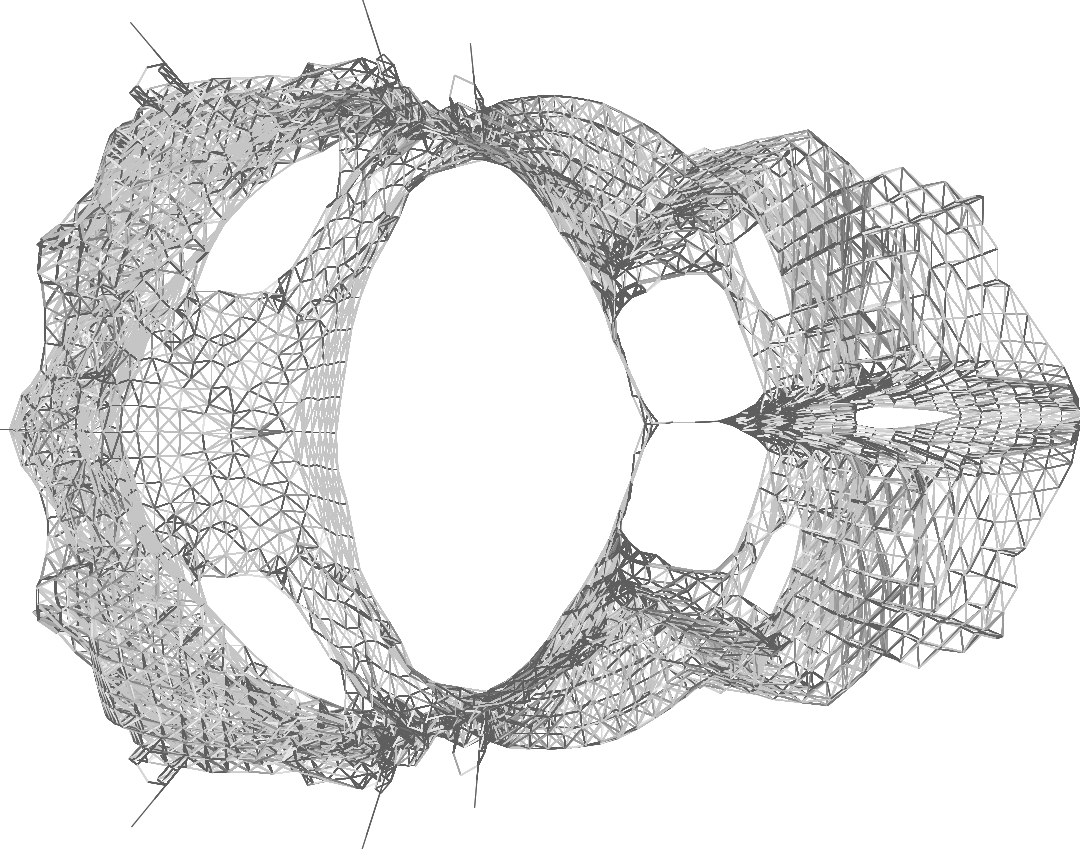}  
\hfill
\includegraphics[width=4.5cm]{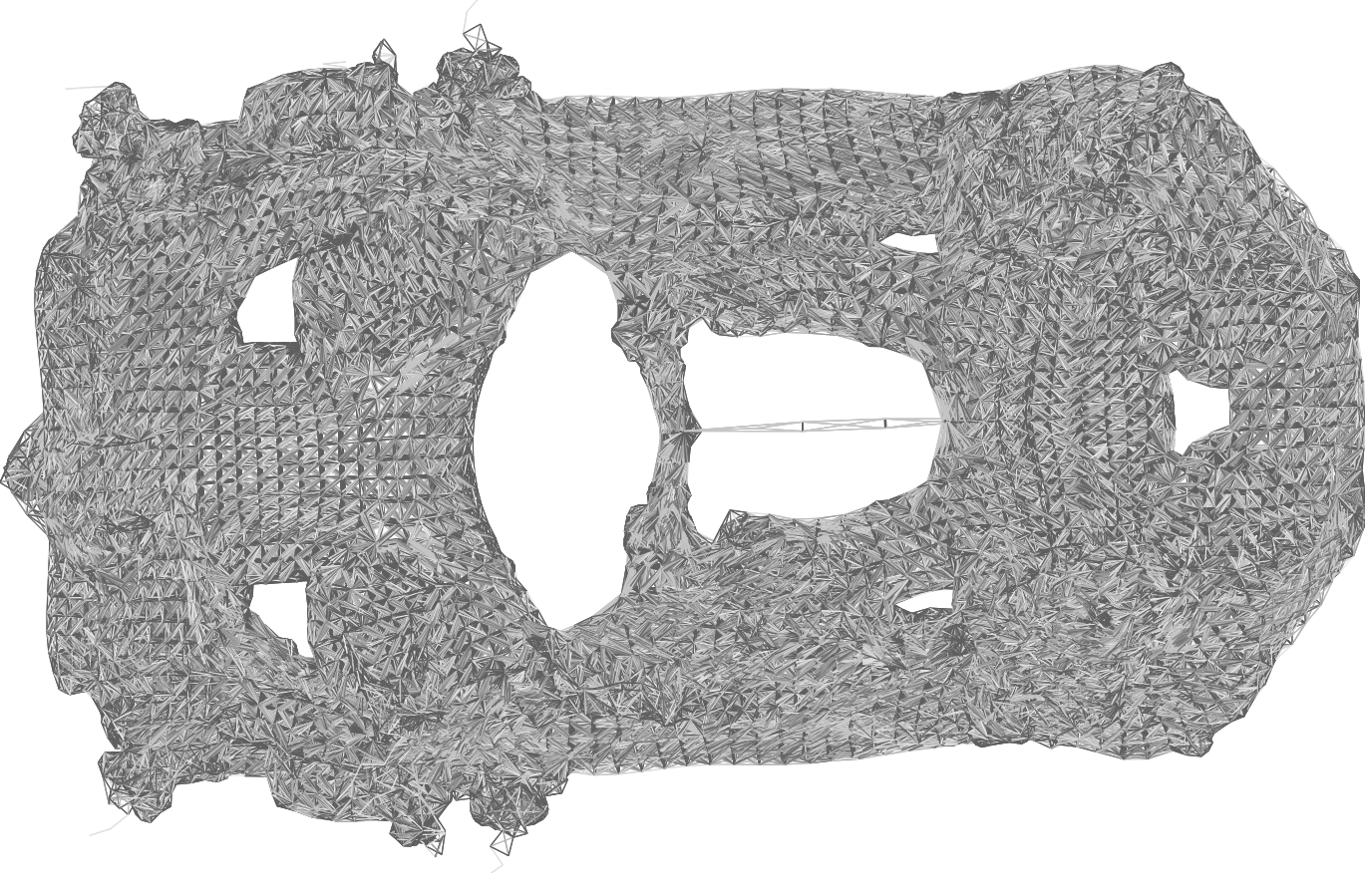}  
\hfill
\includegraphics[width=4.45cm]{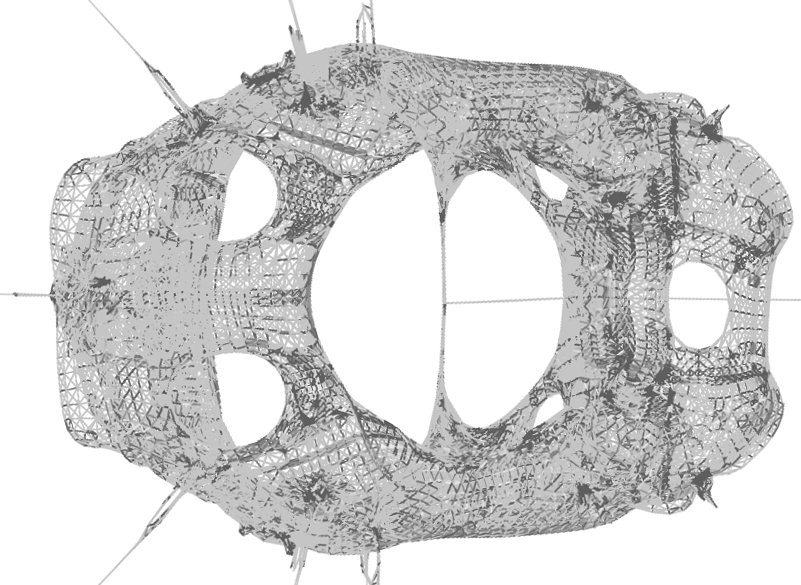}

\caption{Drawings of bcsstk31. Left to right: PivotMDS~\cite{bp-empmsld-07}, GHN~\cite{ghn-mmgl-13}, MulMent (new).}
\vspace*{-.5cm}
\label{fig:drawingsbyalgorithms}
\end{figure*}

Our experimental results show that force approximation rarely
affects the layout quality significantly -- in terms of maxent-stress values as well as visual quality, 
see also Figure~\ref{fig:drawingsbyalgorithms}. The parallel implementation of our multilevel algorithm
 (released as the open source package KaDraw~\cite{kadrawWebsite} -- Karlsruhe Graph Drawing)  with force approximation is, however, on average 30 times faster than the reference 
implementation~\cite{ghn-mmgl-13} -- and even our sequential approximate algorithm is faster than the reference.
A contribution besides higher speed is that, in contrast to~\cite{ghn-mmgl-13}, our approach does not require
input coordinates to optimize the maxent-stress measure.

\chapter{Practical Kernelization}
There are many real-world optimization problems can be formulated as a graph problem. 
Numerous of these problems are NP-hard: it is expected that no efficient (polynomial-time) algorithm exists that always finds an optimal solution. 
However, many NP-hard graph problems have been shown to be fixed-parameter tractable (FPT): large inputs can be solved efficiently and provably optimally, as long as some problem parameter is small. 
Here the parameter measures the `difficulty' of the input in some mathematically well-defined way, for example using the treewidth of the underlying graph. 
Over the last two decades, significant advances have been made in the design and analysis of FPT algorithms for a wide variety of graph problems. 
This has resulted in a rich algorithmic toolbox that are by now well-established and are described in several textbooks and surveys. They lead to algorithms that are theoretically efficient: they allow problems of size $n$ with a parameter value of $k$ to be solved in time $f(k)n^c$ for some (exponential) function $f$ and constant $c$, thereby restricting the exponential dependence of the running time to the parameter $k$ only.
However, these theoretical algorithmic ideas have received very little attention from the practical perspective. Few of the new techniques are implemented and tested on real datasets, and their practical potential is far from understood. The rich toolbox of parameterized algorithm theory offers a rich set of algorithmic ideas that are challenging to implement and engineer in practical settings. By applying techniques from FPT algorithms in nontrivial ways, algorithms can be obtained that perform surprisingly well on real-world instances for NP-hard problems. 

In this line of research, we started to investigate two problems in this context: the independent set problem \cite{evoIS,kerMIS,MISonthefly,misatscalejv,parallelmis,exactWMIS} and the minimum cut problem~\cite{mincut,mincutjv,smexact}.
In general, we follow the scheme of iteratively applying exact and inexact reduction/kernelization rules to compute a much smaller problem kernel on which the problem under consideration is solved.
The distinction between exact and inexact here is as follows: if we have an optimum solution on the problem kernel we can construct an optimum solution on the input problem when only using exact reduction rules. This is not the case for inexact reduction rules. 
In our research, the problem kernel is typically solved using an exact algorithm or an heuristic algorithm.

\section{Independent Sets}
\label{sec:independentset}
\mycoauthor{This section is based on \cite{evoIS,kerMIS,MISonthefly,misatscalejv,parallelmis,exactWMIS} which are joint publications with Jakob Dahlum, Demian Hespe, Sebastian Lamm, Peter Sanders, Darren Strash, Renato F. Werneck, Robert Williger and Huashuo Zhang.}
The maximum (weight) independent set problem is an NP-hard problem that has attracted much attention in the combinatorial optimization community, due to its difficulty and its importance in many fields.
Given a graph $G=(V,E,w)$ with weight function $w:V\rightarrow \mathbb{R}^+$, the goal of the maximum weight independent set problem is to compute a set of vertices $\mathcal{I}\subseteq V$ with maximum total weight, such that no vertices in $\mathcal{I}$ are adjacent to one another.
Such a set is called a \emph{maximum weight independent set} (MWIS). The problem is called maximum independent set problem if one seeks to find the independent set with the largest cardinality, \ie the weight function is identical to one.
The maximum (weight) independent set problem has applications spanning many disciplines, including signal transmission, information retrieval, and computer vision~\cite{balas1986finding}.
As a concrete example, weighted independent sets are vital in labeling strategies for maps~\cite{gemsa2014dynamiclabel,barth-2016},
where the objective is to maximize the number of visible non-overlapping labels on a map.
Here, the maximum weight independent set problem is solved in the label conflict graph, where any two overlapping labels are connected by an edge and vertices have a weight proportional to the city's population.

Similar to their unweighted counterparts, a maximum weight independent set $\mathcal{I}\subseteq V$ in $G$ is a maximum weight clique in $\overline{G}$ (the complement of $G$), and $V \setminus \mathcal{I}$ is a minimum vertex cover of $G$~\cite{xu2016new,cai-dynwvc}.
Since all of these problems are NP-hard~\cite{garey1974}, heuristic algorithms are often used in practice to efficiently compute solutions of high quality on \emph{large} graphs~\cite{pullan-2009,hybrid-ils-2018,li2017efficient,cai-dynwvc}.

We started to investigate the independent set problem by engineering an advanced memetic algorithm~\cite{evoIS}.
The core innovations of the algorithm are very natural combine operations based on graph partitioning and local search algorithms.  More precisely, we employ a state-of-the-art graph partitioner to derive operations that enable us to quickly exchange whole blocks of given independent sets. To enhance newly computed offsprings we combine our operators with a local search algorithm (see Section~\ref{evolutionary} for more details). 
Since evolutionary computation can take some time, we combined this with kernelization techniques in \cite{kerMIS,misatscalejv}. The kernelization techniques have previously been used in branch-and-reduce frameworks to solve the problem to optimality.
More precisely, a recent exact algorithm by Akiba and Iwata~\cite{akiba-2015} has shown that large networks can be solved exactly by employing a branch-and-reduce technique that recursively kernelizes the graph and performs branching. However, one major drawback of their algorithm is that, for huge graphs, branching still can take exponential time.
The key idea of our approach is to apply the memetic algorithm on the problem kernel.
Our algorithm may be viewed as performing two functions simultaneously: (1) reduction rules are used to boost the performance of the memetic algorithm \emph{and} (2) the memetic algorithm opens up the opportunity for further reductions by selecting vertices that are likely to be in large independent sets. In short, our method applies reduction rules to form a kernel, then computes vertices to insert into the final solution and removes their neighborhood (including the vertices themselves) from the graph so that further reductions can be applied. This process is repeated recursively, discovering large independent sets as the recursion proceeds. We show that this technique finds large independent sets much faster than existing local search algorithms, is competitive with state-of-the-art exact algorithms for smaller graphs, and allows us to compute large independent sets on huge sparse graphs, with billions of edges.
In addition, our new algorithm computes an optimal independent set on all instances that the exact algorithm can solve.  

A major drawback of these algorithms is that they require significant preprocessing overhead, and therefore cannot be used to find a high-quality independent set quickly.
We show in \cite{MISonthefly} that performing simple kernelization techniques in an online fashion significantly boosts the performance of local search, and is much faster than pre-computing a kernel using advanced techniques. More precisely, for some reduction rules we can just mark vertices as removed and the local search algorithm does not have to consider these vertices anymore. In addition, we show that cutting high-degree vertices can boost local search performance even further, especially on huge (sparse) complex networks.
Our experiments show that we can drastically speed up the computation of large independent sets compared to other state-of-the-art algorithms, while also producing results that are very close to the best known solutions.

As we noted already, it is of critical importance for these algorithms that kernelization is fast and returns a small kernel.  Hence, we give an efficient parallel kernelization algorithm based on graph partitioning and parallel bipartite maximum matching~\cite{parallelmis}. We combine our parallelization techniques with two techniques to accelerate kernelization further: dependency checking that prunes reductions when they will provably not succeed, therefore significantly reducing the number of failed reduction, and reduction tracking that allows us to stop kernelization when reductions become less fruitful.  
Note, however, that this is a trade-off between size of the kernel and kernelization speed. 
Our algorithm produces kernels that are orders of magnitude smaller than the fastest kernelization methods, while having a similar execution time. 
Furthermore, our algorithm is able to compute kernels with size comparable to the smallest known kernels, but up to two orders of magnitude faster than previously possible.
The project has been released as the KaMIS~\cite{kamisWebsite} (Karlsruhe Maximum Independent Sets) open source framework.

Lastly, we investigate the weighted independent set problem \cite{exactWMIS}.
While kernelziation is known to be very effective in practice for the unweighted version of the problem, very little is known for weighted problem, in part due to a lack of known effective reductions. 
We develop a full suite of new reductions for the maximum weight independent set problem and provide extensive experiments to show their effectiveness in practice on real-world graphs of up to millions of vertices and edges.
Our experiments indicate that our approach is able to far outperform existing state-of-the-art algorithms, solving many instances that were previously infeasible. In particular, we show that branch-and-reduce is able to solve a large number of real-world instances up to two orders of magnitude faster than existing (inexact) local search algorithms---and is able to solve the majority of instances within 15 minutes. For those instances remaining infeasible, we show that combining kernelization with local search produces higher-quality solutions than local search alone.

\section{Minimum Cuts}
\mycoauthor{This section is based on \cite{mincut,mincutjv,smexact} which are joint publications with Monika Henzinger, Alexander Noe and Darren Strash.}
The second problem that we tackled in this line of research is the minimum cut problem.
Given an undirected graph with non-negative edge weights,
the \emph{minimum cut problem} is to partition the vertices into two sets so
that the sum of edge weights between the two sets is minimized. A minimum cut is
often also referred to as the \textit{edge connectivity} of a
graph~\cite{nagamochi1992computing,henzinger2017local}. The problem has
applications in many fields. In particular, for network
reliability~\cite{karger2001randomized,ramanathan1987counting}, 
assuming equal failure chance on edges, the smallest edge cut in the network has
the highest chance to disconnect the network; in VLSI
design~\cite{krishnamurthy1984improved}, a minimum cut can be used to minimize the number of
connections between microprocessor blocks; and it is further used as a subproblem in the
branch-and-cut algorithm for solving the Traveling Salesman Problem and other
combinatorial problems~\cite{padberg1991branch}.

For minimum cut algorithms to be viable for these (and other) applications they must be fast on small data sets and scale to large data sets. Thus, an algorithm should have either linear or near-linear running time, or have an efficient parallelization. Note that \emph{all} existing exact algorithms have non-linear running time~\cite{hao1992faster,henzinger2017local,karger1996new}, where the fastest of these is the deterministic algorithm of Henzinger~\etal\cite{henzinger2017local} with running time $\Oh{m \log^2{n} \log \log^2 n}$. Although this is arguably near-linear theoretical running time, it is not known how the algorithm performs in practice. Even the randomized algorithm of Karger and Stein~\cite{karger1996new} which finds a minimum cut only with high probability, has $\Oh{n^2\log^3{n}}$ running time, although this was later improved by Karger~\cite{karger2000minimum} to $\Oh{m\log^3{n}}$. There is a linear time approximation algorithm, namely the $(2 + \varepsilon)$-approximation algorithm by Matula~\cite{matula1993linear}. However, the quality of Matula's algorithm in practice is currently unknown---no experiments have been published, although Chekuri \etal provide an implementation~\cite{Chekuri:1997:ESM:314161.314315,code}.

We to give the first \emph{practical} shared-memory parallel algorithm for the minimum cut problem~\cite{mincut,mincutjv}. Our first algorithm is heuristic (\ie it does not give guarantees on solution quality), randomized, and has running time $\Oh{n+m}$ when run sequentially. The algorithm works in a multilevel fashion: we repeatedly reduce the input graph size with both heuristic and exact techniques, and then solve the smaller remaining problem with exact methods.
Our heuristic technique identifies edges that are unlikely to be in a minimum cut using label propagation
introduced by Raghavan~\etal\cite{raghavan2007near} and contracts them in bulk.
We further combine this technique with the exact reduction routines
from Padberg and Rinaldi~\cite{padberg1990efficient}. 
For example, given a bound on the minimum cut $\hat\lambda$, it is very obvious that one can contract every edge having weight larger than $\hat\lambda$ without loosing optimality.
We perform extensive experiments comparing our algorithm with other heuristic algorithms as well as exact algorithms on real-world and generated instances, which include graphs on up to \numprint{70} million vertices and \numprint{5} billion edges---the largest graphs ever used for experiments for the minimum cut problem.
Results indicate that our algorithm finds optimal cuts
on almost all instances and also that the empirically observed error rate is lower than competing
heuristic algorithms that come with guarantees on the solution quality. At the same time, even when run sequentially, our algorithm is significantly faster (up to a factor of $4.85$) than other state-of-the-art algorithms.

Inspired by the good performance of this algorithm and the low error rate, we engineer the fastest known \emph{exact} minimum cut algorithm for the problem~\cite{smexact}. 
We do so by (1) incorporating the proposed inexact methods and (2) by using better suited data structures and other optimizations as well as (3) parallelization of exact methods.
Our algorithm achieves improvements in running time by a multitude of techniques. First, we use the fast and parallel \emph{inexact} minimum cut algorithm from above~\cite{mincut,mincutjv} to obtain a better approximate bound $\hat\lambda$ for the problem (recall that the algorithm almost always gave the correct result). As known reduction techniques depend on this bound, the better bound enables us to apply more reductions and to reduce the size of the graph much faster.
For example, edges whose incident vertices have a connectivity of at least $\hat\lambda$, can be contracted without the contraction affecting the minimum cut.
Nagamochi \etal\cite{nagamochi1992computing,nagamochi1994implementing} give reductions that use maximum spanning forests to find a non-empty set of contractible edges.
The intuition behind the algorithm is as follows: imagine you have an unweighted graph with minimum cut  value exactly one. 
Then any spanning tree must contain at least one edge of each of the minimum cuts. 
Hence, after computing a spanning tree, every remaining edge can be contracted without losing the minimum cut.
Nagamochi, Ono and Ibaraki extend this idea to the case where the graph can have edges with positive weight as well as the case in which the minimum cut is bounded by $\hat \lambda$ and show how edges are identified using one modified breadth first search.

Using better suited data structures as well as incorporating observations that help to save a significantly amount of work in the contraction routine of Nagamochi, Ono and Ibaraki~\cite{nagamochi1994implementing} further reduce the running time of our algorithm. Additionally, we give a parallel variant of the contraction routines of Nagamochi, Ono and Ibaraki~\cite{nagamochi1994implementing}.
We arrive at an exact algorithm (released as the open source package VieCut~\cite{viecutWebsite} -- Vienna Minimum Cuts) that outperforms the state-of-the-art by a factor of up to $2.5$ already sequentially, and when run in parallel by a factor of up to $12.9$ using~$12$~cores.

\chapter{Parallelization}

Complex graphs, which are useful in a wide range of applications, can sometimes be composed out of billions of entities that give rise to emerging properties and structures.
Analyzing these structures aids us in gaining new insights about our surroundings.
With the recent stagnation of Moore's law, the primary method for gaining computing power is to increase the number of available cores, processors, or networked machines (all of which are generally referred to as \emph{processing elements} (PEs)) and exploit parallel computation. 
Designing and evaluating algorithms to handle these datasets is a crucial task on the road to understanding the underlying systems.
Hence, scalable shared-memory and distributed-memory parallel graph algorithms for parallel analysis that efficiently utilize all cores of a machine or use many machines of a super computer are highly desirable.

For example, graph partitioning (GP) is a key prerequisite for efficient large-scale parallel graph algorithms. 
In many cases, a graph needs to be partitioned or clustered such that there are few edges between the blocks (pieces).  
In particular, when you process a graph in parallel on $k$ PEs (processing elements), you often want to partition the graph into $k$ blocks of about equal
size.  In this paper we focus on a version of the problem that constrains the
maximum block size to $(1+\epsilon)$ times the average block size and tries to
minimize the total cut size, i.e., the number of edges that run between blocks.
In Section~\ref{multilevelpart}, we have already seen our parallel approaches to graph partitioning~\cite{externalcomplexnetworks,parallelsnw,sharedmempart,parallelsnwjv}.

A lot more of the graph algorithms already proposed in this work fall into the parallelization category. In graph drawing~\cite{drawing,jvdrawing}, we used shared-memory parallelism to parallelize local optimization of drawings. We propose parallel reductions for the independent set problem~\cite{parallelmis} where we our parallel graph partitioning algorithm and then employed one thread per block to run reductions locally. Lastly, we proposed shared-memory parallel minimum cut algorithms~\cite{mincut,mincutjv,smexact}. Here we used a shared-memory parallel label propagation algorithms as well as parallel reduction routines. In Section~\ref{evolutionary}, we propose parallel memetic algorithms. We now extend this to distributed-memory parallel algorithms that do not directly fit into the other research pillars, \ie parallel graph generation~\cite{bagenjv,graphgennew,jvgraphgennew} and distributed edge partitioning algorithms~\cite{scalableedgepartitioning}. 
\section{Graph Generation}
\mycoauthor{This section is based on \cite{bagenjv,graphgennew,jvgraphgennew} which are joint publications with Daniel Funke, Sebastian Lamm, Ulrich Meyer, Peter Sanders, Darren Strash and Moritz von Looz.}
Building scalable graph algorithms is a challenging task that requires a careful analysis and an extensive evaluation.
However, engineering such algorithms is often hindered by the scarcity of publicly~available~datasets.
Network generators serve as a tool to alleviate this problem by providing synthetic instances with controllable parameters.
However, many network generators fail to provide instances on a massive scale due to their sequential nature or resource constraints.
Additionally, truly scalable network generators are few and often limited in their realism.

Scale-free graphs with a power-law degree distribution seem to be
ubiquitous in complex network analysis. In order to study such
networks and the algorithms to analyze them, one needs simple models for
generating complex networks with user-definable parameters.
In their
seminal paper \cite{BA} Barabasi and Albert define the model that is
perhaps most widely used because of its simplicity and intuitive
definition: We start with an arbitrary seed network consisting of
nodes $0..n_0-1$ ($a..b$ is used as a shorthand for $\set{a,\ldots,b}$
here).  Nodes $i\in n_0..n-1$ are added one at a time. They randomly
connect to $d$ neighbors using \emph{preferential attachment}, i.e.,
the probability to connect to node $j\leq i$ is chosen proportionally
to the degree of $j$. The seed graph, $n_0$, $d$, and $n$ are
parameters defining the graph family.

We started this line of research by developing a scalable graph generator for the Barabasi-Albert model~\cite{bagenjv}.
Our starting point is the fast, simple, and elegant sequential
algorithm by Batagelj and Brandes \cite{Brandes}.  For simplicity of
exposition, we use an empty seed graph ($n_0=0$).  A generalization
only requires a number of straight forward index transformations.
Brandes' algorithm generates one edge at a time and writes it into an
edge array $E[0..2dn-1]$ where positions $2i$ and $2i+1$ store the node
IDs of the end points of edge $i$.  We have $E[2i]=\floor{i/d}$.  The central
observation is that one gets the right probability distribution for the other end point by
uniformly sampling edges rather than sampling dynamically weighted
nodes, \ie $E[2i+1]$ is simply set to $E[x]$ where $x$ is chosen
uniformly and (pseudo)randomly from $0..2i$.

The idea behind the parallel algorithm~\cite{bagenjv} is very simple -- compute edge
$i$ independently of all other edges and without even accessing the
array $E$. On the first glance, this sounds paradoxical because there
\emph{are} dependencies and accessing $E$ is the whole point behind
Brandes' algorithm.  This paradox is resolved by the idea to 
recompute any entry of $E$ that is needed for edge $i$ using hash functions as pseudorandom number generator. 

Hence, our parallel algorithm \emph{requires no communication at all} and yields perfect load balance on uniform nodes of a supercomputer. Hence, our algorithm achieves perfect scalability.
We generated a Petaedge graph in less than an hour on 16\,384 cores of the SuperMUC computer.
This graph is 20\,000 times larger than the largest Barabasi-Albert graph we have seen reported.

Based on the idea to have distributed parallel algorithms that do not need to communicate~\cite{graphgennew,jvgraphgennew}, we started to build more graph generators for a variety of network models commonly found in practice.
The models that we investigated are the classic \erdos~models $G(n,m)$ and $G(n,p)$ and different spatial network models including random geometric graphs (RGGs), random hyperbolic graphs (RHGs) and random Delaunay graphs (RDGs).
For each generator, we provide bounds for their parallel (and sequential) running times.
A key-component of our algorithms is the clever use of pseudorandomization and divide-and-conquer strategies.
These components enable us to perform efficient recomputations in a distributed setting without the need for communication. 

To highlight the practical impact of our generators, we also present an extensive experimental evaluation.
First, we show that our generators rival the current state-of-the-art in terms of sequential and/or parallel running time.
Second, we are able to show that our generators have near optimal scaling behavior in terms of weak scaling (and strong scaling).
Finally, our experiments show that we are able to produce instances of up to $2^{43}$ vertices and $2^{47}$ edges in less than 22 minutes.
These instances are in the same order of magnitude as those generated by R-MAT for the Graph~500 benchmark~(\url{http://www.graph500.org}).
Hence, our generators enable the underlying network models to be used in massively distributed settings.
The generators are available in the open source KaGen~\cite{kagenWebsite} -- Karlsruhe Graph Generation -- package. 

\section{Edge Partitioning}
\mycoauthor{This section is based on \cite{scalableedgepartitioning} which is joint work Sebastian Schlag, Daniel Seemaier and Darren Strash.}
As already mentioned, one useful method to take advantage of parallelism is found in \emph{graph partitioning} which attempts to partition the vertices of a graph into roughly equal disjoint sets (called \emph{blocks}), while minimizing some objective function---for example minimizing the number of edges crossing between blocks. 
This traditional (node-based) graph partitioning has also been essential for making efficient distributed graph algorithms in the Think Like a Vertex (TLAV) model of computation~\cite{mccune-tlav-2015}. In this model, node-centric operations are performed in parallel, by mapping nodes to PEs and executing node computations in parallel. Nearly all algorithms in this model require information to be communicated between neighbors --- which results in network communication if stored on different PEs --- and therefore high-quality graph partitioning directly translates into less communication and faster overall running time. 
However, node-centric computations have serious shortcomings on power law graphs --- which have a skewed degree distribution. In such networks, the overall running time is negatively affected by very high-degree nodes, which can result in more communication steps. To combat these effects, Gonzalez~\etal\cite{gonzalez-powergraph-2012} introduced edge-centric computations, which duplicates node-centric computations across edges to reduce communication overhead. In this model, \emph{edge partitioning}---partitioning edges into roughly equally sized blocks---must be used to reduce the overall running time. 

Similar to (node-based) graph partitioning, the quality of the edge partitioning can have a dramatic effect on parallelization~\cite{li2017spac,Patwary:2010}.
Noting that edge partitioning can be solved directly with hypergraph partitioners Li~\etal\cite{li2017spac} showed that these techniques give the highest quality partitionings; however, they are also slow.
Therefore, a balance of solution quality and speed must be taken into consideration. This balance is struck well for
the split-and-connect (SPAC) method introduced by Li~\etal\cite{li2017spac}. In the SPAC method, vertices are duplicated and weighted so that a (typically fast) standard node-based graph partitioner can be used to compute an edge partitioning; however, this method was only studied in the sequential setting. 
However, distributed algorithms for the problem fare far worse~\cite{gonzalez-powergraph-2012,bourse-2014}. While adding much computational power with many processing elements (PEs), edge partitioners such as  PowerGraph~\cite{gonzalez-powergraph-2012} and Ja-Be-Ja-VC~\cite{fatemeh2014jabejavc}, produce partitionings of significantly worse quality than those produced with hypergraph partitioners or SPAC. Thus, there is no clear winning algorithm.

We give the first \emph{high-quality} distributed-memory parallel edge partitioner~\cite{scalableedgepartitioning}. Our algorithm scales to networks with billions of edges, and runs efficiently on thousands of PEs.
Our technique is based on a fast parallelization of split-and-connect graph construction and a use of advanced 
node partitioning algorithms. 

More precisely, given an undirected, unweighted graph $G = (V, E)$, the split-and-connect graph $G' = (V', E', c', \omega')$ is constructed as follows: 
for each node $v \in V$, create a set of \emph{split nodes} $S_v := \{v'_1, \dots, v'_{d(v)}\}$ that are connected to a cycle by \emph{auxiliary edges} with edge-weight one, \ie edges $\{v'_i, v'_{i + 1}\}$ for $i = 1, \dots, d(v) - 1$ and $\{v'_{d(v)}, v'_1\}$. 
In the connect phase, split nodes are connected by edges, \ie for each edge $e = \{u,v\}$ in $G$, a corresponding \emph{dominant edge} $\{u',v'\}$ in $G'$ is created.
This is done such that overall both $u' \in S_u$ and $v' \in S_v$ are connected to one and only one dominant edge.
Those dominant edges get assigned edge weight infinity.
To partition the edges of $G$, a (parallel) node-based partitioning algorithm is run on $G'$. 
Since the vertex cut is always smaller than or equal to the edge cut, a good node partition of the split-and-connect graph intuitively leads to a good edge partition of the input graph. 

Our experiments show that while hypergraph partitioners outperform SPAC-based graph partitioners in the sequential setting regarding both solution quality and running time,
our new algorithms compute significantly \emph{better} solutions than the distributed-memory hypergraph partitioner Zoltan~\cite{Zoltan} in \emph{shorter} time.
Moreover, our techniques scale well to \numprint{2560} PEs, allowing for efficient partitioning of graphs with billions of edges within seconds. The techniques are integrated in the KaHIP framework.

\chapter{Memetic Algorithms}
\label{evolutionary}

The last pillar of this work is concerned with memetic algorithms. We develop several memetic algorithms, i.e.~for the node separator problem~\cite{evoNS}, the hypergraph partitioning problem~\cite{evoHG}, the DAG partitioning problem~\cite{evoDAG}, the territory design problem~\cite{areaDesignGIS}, as well as the graph clustering problem~\cite{clustering}.
For most of these problems, we had already developed multilevel algorithms to tackle the problems (see Chapter~\ref{sec:multilevel}).
However, we also developed a memetic algorithms, for the independent set problem, that are not per se based on the multilevel paradigm~\cite{evoIS,kerMIS,misatscalejv}.

\section{Multilevel-based Memetic Algorithms}
\mycoauthor{This section is based on \cite{evoHG,evoDAG,areaDesignGIS,evoNS,clustering} which is joint work Nitin Ahuja, Robin Andre, Matthias Bender, Sonja Biedermann, Monika Henzinger, Orlando Moreira, Merten Popp, Peter Sanders, Sebastian Schlag, Darren Strash, Andreas Wagner, Robert Williger.}
Recall, the intuition of the multilevel scheme is that a good solution at one level of the hierarchy will also be a good solution on the next finer level. Hence, 
depending on the definition of the neighborhood, local search algorithms are able to explore local solution spaces very effectively~in~this~setting. 
However, these methods are also prone to get trapped in local optima. 
The multilevel scheme can help to some extent since 
local search has a more global view on the problem on the coarse levels and a very fine-grained view on the fine levels of the multilevel hierarchy. 
In addition, as with many other randomized meta-heuristics, several repeated runs can be made in order to improve the final result at the expense of running time.
Still, even a large number of repeated executions can only scratch the surface of the huge search space of possible clusterings. 
In order to explore the global solution space extensively, we need more sophisticated meta-heuristics. 
This is where memetic algorithms (MAs), i.e.~genetic algorithms combined with local search, 
come into play. \emph{Memetic algorithms} allow for effective exploration (global search) and exploitation (local search) of the solution space.

All of the memetic algorithms in this section~\cite{evoHG,evoDAG,areaDesignGIS,evoNS,clustering} go hand in hand with the multilevel scheme, \ie the scheme yields us a new effective way of combining different solutions.
Our algorithms use the following simple evolution scheme:
we start with a population of individuals (this depends on the problem under consideration) and evolves the population into different populations over several rounds. 
In each round, the EA uses a selection rule based on the fitness of the individuals (depends on the objective of the considered problem) of the population to select good individuals and combine them to obtain improved offspring. 
When an offspring is generated an elimination rule is used to select a member of the population and replace it with the new offspring. 

The core-innovations of our memetic algorithms are the recombination operators that make use of the multilevel scheme of the respective problem.
Typically, our recombination operators ensure that the offspring has an objective \textit{at least as good as the best of both parents}.  
Roughly speaking, the recombination operator does not contract any cut (hyper-)edge of both inputs and is thus able to use the better of both inputs solutions on the coarsest level of a multilevel scheme. 
To give an example, we focus on the graph clustering case~\cite{clustering}: Let $\mathcal{C}_1$ and $\mathcal{C}_2$ be two individuals from the population (which in this case are two different clusterings). 
Both individuals are used as input for the multilevel method in the following sense. 
Let $\mathcal{E}$ be the set of edges run between two blocks of the clusterings in either $\mathcal{C}_1$ \emph{or} $\mathcal{C}_2$. 
All edges in $\mathcal{E}$ are blocked during the coarsening phase, \ie they are \emph{not} contracted during the coarsening phase.
In other words, these edges cannot be contracted during the multilevel scheme. 
We stop contracting clusterings when no contractable edge is left. 
Depending on the problem at hand, the definition has to be adapted. For example, in the hypergraph partitioning case~\cite{evoHG} we contract any two pairs that are from the same block and in the node separator case~\cite{evoNS}, we disallow edges leaving any of the input separators for contraction.

As typical local search algorithm ensure that solution quality is not worsened, the output can potentially be improved during the uncoarsening phase of the multilevel scheme.
This is true for all problems mentioned in this section except the territory design problem (which uses a different objective function).
Note however that although the general scheme is in every case very similar, problem specific adaptions have to be made in order to make the recombination operations work for any of the problems.

For DAG partitioning~\cite{evoDAG}, territory design~\cite{areaDesignGIS}, the node separator problem~\cite{evoNS} as well as the graph clustering problem~\cite{clustering}, we employ additional coarse-grained parallelization.
We now explain the island-based parallelization that we use. 
Each processing element (PE)  basically performs the same operations using different random seeds.
First, we estimate the population size~$\mathcal{S}$: each PE creates an individual and measures the time $\overline{t}$ spent. 
We then choose $\mathcal{S}$ such that the time for creating $\mathcal{S}$ clusterings is approximately $t_{\text{total}}/c$ where the fraction $c$ is a tuning parameter and $t_{\text{total}}$ is the total running time that the algorithm is given to produce a clustering of the graph. 
The minimum amount of individuals in the population is set to~3, the maximum amount of the individuals in the population is set to 100. 
The lower bound on the population size is chosen to ensure a certain minimum of diversity, while
the upper bound is used to ensure convergence. 

Each PE then builds its own population. 
Afterwards, the algorithm proceeds in rounds as long as time is left. 
Either a mutation or recombination operation is performed. 
Our communication protocol is similar to \textit{randomized rumor spreading} which has shown to be scalable in previous work~\cite{kaffpaE}. 
Let $p$ denote the number of PEs used. A communication step is organized in rounds. 
In each round, a PE chooses a communication partner and sends her the currently best individual $\mathcal{C}$ of the local population. 
The selection of the communication partner is done uniformly at random among those PEs to which $\mathcal{C}$ not already has been sent to.  
Afterwards, a PE checks if there are incoming individuals and if so inserts them into the local population using the elimination strategy described above.
If the best local individual has changed, all PEs are again eligible.
This is repeated $\log p$ times.
The algorithm is implemented \textit{completely asynchronously}, \ie there is no need for a global synchronization.

We typically arrive at systems that are more effective than repeated executions of the multilevel scheme and produce solutions with high solution quality.
For example, the clustering system~\cite{clustering} is able to reproduce or improve previous all entries of the 10th DIMACS implementation challenge under consideration as well as results recently reported in the literature in a short amount of time.
Moreover, while the previous best result for different instances has been computed by a variety of solvers, our algorithm can now be used as a single tool to compute the result.
Hence, overall we consider our algorithm as a new state-of-the-art heuristic for solving the modularity clustering problem. The algorithms can be obtained in the VieClus~\cite{vieclusWebsite} -- Vienna Graph Clustering -- open source package.

As another example, experiments indicate that our memetic hypergraph partitioning algorithm~\cite{evoHG} is able to compute partitions of very high quality, scales well to large networks,
and performs better than KaHyPar, which seems to be the current method of choice among the available hypergraph partitioning tools unless speed is more important than quality~\cite{hs2017sea}.
In a setting where competing algorithms get the same fairly large amount of time to compute a solution, 
   our new algorithm computes the best result on $597$ out of the $630$  benchmark instances. 
This is in contrast to previous \emph{non-multilevel} evolutionary algorithms for the problem which are not considered to be competitive with state-of-the-art tools~\cite{Cohoon2003}.  

\vfill\pagebreak
\section{Other Memetic Algorithms}
\mycoauthor{This section is based on \cite{evoIS,kerMIS,misatscalejv} which is joint work Sebastian Lamm, Peter Sanders, Darren Strash, and Renato F.\@ Werneck.}

We also developed memetic algorithms for the independent set problem~\cite{evoIS,kerMIS,misatscalejv}.
The core innovations of the algorithm are recombine operations based on graph partitioning and local search algorithms~\cite{evoIS}.
More precisely, we employ a graph partitioner to derive operations that enable us to quickly \emph{exchange whole blocks} of given individuals.  
We explain the most simple version of the recombine operations: 
In its simplest form, the operator starts by computing a node separator $V=V_1 \cup V_2 \cup S$ of the input graph (using our graph partitioning algorithms). 
The node separator has the property that there are no edges running between $V_1$ and $V_2$.
We then use the separator $S$ as a crossover point for our operation.
The operator generates two offsprings from two input independent sets $\mathcal{I}_1, \mathcal{I}_2$ (individuals from the population).
More precisely, we set $O_1=(V_1\cap \mathcal{I}_1) \cup (V_2\cap\mathcal{I}_2)$ and $O_2=(V_1\cap \mathcal{I}_2) \cup (V_2\cap\mathcal{I}_1)$. 
In other words, we exchange whole parts of independent sets from the blocks $V_1$ and $V_2$ of the node separator. 
Note that the exchange can be implemented in time linear in the number of nodes.
Hence, the computed offsprings are independent sets, but may not be maximal since separator nodes have been ignored and potentially some of them can be added to the solution.
We maximize the offsprings by using local search. 
In \emph{contrast} to previous evolutionary algorithms, each computed offspring is valid.
Hence, we only allow valid solutions in our population and thus are able to use the cardinality of the independent set as a fitness function.
Since evolutionary computation can take some time, we combined this with kernelization techniques in \cite{kerMIS,misatscalejv}. The kernelization techniques have previously been used in branch-and-reduce frameworks to solve the problem to optimality.
The key idea of our approach is to apply the memetic algorithm on the problem kernel (see Section~\ref{sec:independentset} for more details).
Experiments indicate that our algorithms outperform state-of-the-art algorithms on large variety of instances -- some of which are better than every reported in literature.

\chapter{Funding}

There have been multiple grants from which this research has received funding. While the grants are acknowledged in the individual papers, we also acknowledge them here.
The research leading to these results has received funding from the European Research Council under the European Union's Seventh Framework Programme (FP/2007-2013)~/~ERC Grant Agreement no.~340506. Moreover, this research has been supported by DFG grants SCHU 2567/1-2, the DFG Gottfried Wilhelm Leibniz Prize 2012 for Peter Sanders, by DFG grants SA 933/10-2. Lastly, this research was supported by the Austrian Science Foundation~(FWF, project P 31763-N31) 
\renewcommand\refname{}

\vfill\pagebreak
{
\renewcommand\chapter{}
\renewcommand\bibname{}
\bibliographystyle{beta}
\bibliography{phdthesiscs}

\newcommand{\etalchar}[1]{$^{#1}$}
\begin{thebibliography}{COJT{\etalchar{+}}11}

\bibitem[AHK98]{MLPart}
C.~J. Alpert, J.-H. Huang, and A.~B. Kahng.
\newblock {Multilevel Circuit Partitioning}.
\newblock {\em IEEE Transactions on Computer-Aided Design of Integrated
  Circuits and Systems}, 17(8):655--667, 1998.

\bibitem[AI15]{akiba-2015}
T.~Akiba and Y.~Iwata.
\newblock {Branch-and-reduce Exponential/FPT Algorithms in Practice: A Case
  Study of Vertex Cover}.
\newblock In {\em Proceedings of the Meeting on Algorithm Engineering \&
  Expermiments}, ALENEX'15, pages 70--81. SIAM, 2015.

\bibitem[AMS04]{DBLP:journals/jucs/AloulMS04}
F.~A. Aloul, I.~L. Markov, and K.~A. Sakallah.
\newblock {{MINCE:} {A} Static Global Variable-Ordering Heuristic for {SAT}
  Search and {BDD} Manipulation}.
\newblock {\em J. {UCS}}, 10(12):1562--1596, 2004.

\bibitem[AvHK06]{abello-ask-06}
J.~Abello, F.~van Ham, and N.~K.
\newblock Ask-graphview: A large scale graph visualization system.
\newblock {\em {IEEE} Transactions on Visualization and Computer Graphics},
  12(5):669--676, 2006.

\bibitem[BA99]{BA}
A.-L. Barabasi and R.~Albert.
\newblock Emergence of scaling in random networks.
\newblock {\em Science}, 286(5439):509--512, 1999.

\bibitem[BB05]{Brandes}
V.~Batagelj and U.~Brandes.
\newblock Efficient generation of large random networks.
\newblock {\em Physical Review E}, 71(3):036113, 2005.

\bibitem[BHSS]{vieclusWebsite}
S.~Biedermann, M.~Henzinger, C.~Schulz, and B.~Schuster.
\newblock {VieClus -- Vienna Graph Clustering}.
\newblock {\url{http://vieclus.taa.univie.ac.at/}}.

\bibitem[BJ92a]{BuiJ92}
T.~N. Bui and C.~Jones.
\newblock {Finding Good Approximate Vertex and Edge Partitions is {N}{P}-Hard}.
\newblock {\em Information Processing Letters}, 42(3):153--159, 1992.

\bibitem[BJ92b]{bui1992finding}
T.~N. Bui and C.~Jones.
\newblock {Finding Good Approximate Vertex and Edge Partitions is NP-hard}.
\newblock {\em Information Processing Letters}, 42(3):153--159, 1992.

\bibitem[BL84]{BHATT1984300}
S.~N. Bhatt and F.~T. Leighton.
\newblock {A framework for solving VLSI graph layout problems}.
\newblock {\em Journal of Computer and System Sciences}, 28(2):300 -- 343,
  1984.

\bibitem[BLV14]{bourse-2014}
F.~Bourse, M.~Lelarge, and M.~Vojnovic.
\newblock {Balanced Graph Edge Partition}.
\newblock In {\em Proceedings of 20th ACM SIGKDD International Conference on
  Knowledge Discovery and Data Mining}, KDD '14, pages 1456--1465. ACM, 2014.

\bibitem[BNNS16]{barth-2016}
L.~Barth, B.~Niedermann, M.~N\"{o}llenburg, and D.~Strash.
\newblock Temporal map labeling: A new unified framework with experiments.
\newblock In {\em Proceedings of the 24th ACM SIGSPATIAL International
  Conference on Advances in Geographic Information Systems}, GIS '16, pages
  23:1--23:10. ACM, 2016.

\bibitem[BP07]{bp-empmsld-07}
U.~Brandes and C.~Pich.
\newblock Eigensolver methods for progressive multidimensional scaling of large
  data.
\newblock In {\em Graph Drawing (GD'06)}, volume 4372 of {\em Lecture Notes in
  Computer Science}, pages 42--53. Springer, 2007.

\bibitem[BY86]{balas1986finding}
E.~Balas and C.~S. Yu.
\newblock Finding a maximum clique in an arbitrary graph.
\newblock {\em SIAM Journal on Computing}, 15(4):1054--1068, 1986.

\bibitem[CA99]{PaToH}
{\"{U}}.~V. Cataly{\"u}rek and C.~Aykanat.
\newblock {Hypergraph-Partitioning-Based Decomposition for Parallel
  Sparse-Matrix Vector Multiplication}.
\newblock {\em IEEE Transactions on Parallel and Distributed Systems},
  10(7):673--693, Jul 1999.

\bibitem[CGK{\etalchar{+}}97]{Chekuri:1997:ESM:314161.314315}
C.~S. Chekuri, A.~V. Goldberg, D.~R. Karger, M.~S. Levine, and C.~Stein.
\newblock Experimental study of minimum cut algorithms.
\newblock In {\em Proceedings of the 8th Annual ACM-SIAM Symposium on Discrete
  Algorithms (SODA '97)}, pages 324--333. SIAM, 1997.

\bibitem[CHLL18]{cai-dynwvc}
S.~Cai, W.~Hou, J.~Lin, and Y.~Li.
\newblock Improving local search for minimum weight vertex cover by dynamic
  strategies.
\newblock In {\em Proceedings of the Twenty-Seventh International Joint
  Conference on Artificial Intelligence ({IJCAI} 2018)}, pages 1412--1418,
  2018.

\bibitem[CJZM10]{Application:DistributedDB}
C.~Curino, E.~Jones, Y.~Zhang, and S.~Madden.
\newblock {Schism: A Workload-driven Approach to Database Replication and
  Partitioning}.
\newblock {\em Proceedings VLDB Endow.}, 3(1-2):48--57, September 2010.

\bibitem[CKL03]{Cohoon2003}
J.~Cohoon, J.~Kairo, and J.~Lienig.
\newblock {\em {Evolutionary Algorithms for the Physical Design of VLSI
  Circuits}}, pages 683--711.
\newblock Springer, 2003.

\bibitem[COJT{\etalchar{+}}11]{costa2011analyzing}
L.~F. Costa, O.~N. Oliveira~Jr, G.~Travieso, F.~A. Rodrigues, P.~R. V.~Boas,
  L.~Antiqueira, M.~P. Viana, and L.~E. C.~Rocha.
\newblock {Analyzing and Modeling Real-World Phenomena with Complex Networks: A
  Survey of Applications}.
\newblock {\em Adv. in Physics}, 60(3):329--412, 2011.

\bibitem[DBH{\etalchar{+}}06]{Zoltan}
K.~D. Devine, E.~G. Boman, R.~T. Heaphy, R.~H. Bisseling, and {\"{U}}.~V.
  Cataly{\"u}rek.
\newblock {Parallel Hypergraph Partitioning for Scientific Computing}.
\newblock In {\em 20th International Conference on Parallel and Distributed
  Processing (IPDPS)}, pages 124--124. IEEE, 2006.

\bibitem[DHM{\etalchar{+}}09]{delling2009high}
D.~Delling, M.~Holzer, K.~M{\"u}ller, F.~Schulz, and D.~Wagner.
\newblock High-performance multi-level routing.
\newblock {\em The Shortest Path Problem: Ninth DIMACS Implementation
  Challenge}, 74:73--92, 2009.

\bibitem[DSW14]{dibbelt2014customizable}
J.~Dibbelt, B.~Strasser, and D.~Wagner.
\newblock Customizable contraction hierarchies.
\newblock In {\em 13th International Symposium on Experimental Algorithms
  (SEA'14)}, pages 271--282. Springer, 2014.

\bibitem[FLS{\etalchar{+}}]{kagenWebsite}
D.~Funke, S.~Lamm, P.~Sanders, C.~Schulz, D.~Strash, and M.~von Looz.
\newblock {KaGen -- Karlsruhe Graph Generation}.
\newblock {\url{https://github.com/sebalamm/KaGen/}}.

\bibitem[Geo73]{george1973nested}
A.~George.
\newblock {Nested Dissection of a Regular Finite Element Mesh}.
\newblock {\em SIAM Journal on Numerical Analysis}, 10(2):345--363, 1973.

\bibitem[GHN13]{ghn-mmgl-13}
E.~R. Gansner, Y.~Hu, and S.~C. North.
\newblock A maxent-stress model for graph layout.
\newblock {\em {IEEE} T. Visualization and Computer Graphics}, 19(6):927--940,
  2013.

\bibitem[GJ02]{garey2002computers}
M.~R. Garey and D.~S. Johnson.
\newblock {\em {Computers and Intractability}}, volume~29.
\newblock WH Freeman \& Co., San Francisco, 2002.

\bibitem[GJS74]{garey1974}
M.~R. Garey, D.~S. Johnson, and L.~Stockmeyer.
\newblock {Some Simplified {N}{P}-Complete Problems}.
\newblock In {\em Proceedings of the 6th ACM Symposium on Theory of Computing},
  STOC '74, pages 47--63. ACM, 1974.

\bibitem[GKN05]{gkn-gdsm-05}
E.~R. Gansner, Y.~Koren, and S.~North.
\newblock Graph drawing by stress majorization.
\newblock In {\em Graph Drawing (GD'04)}, volume 3383 of {\em Lecture Notes in
  Computer Science}, pages 239--250. Springer, 2005.

\bibitem[GLG{\etalchar{+}}12]{gonzalez-powergraph-2012}
J.~E. Gonzalez, Y.~Low, H.~Gu, D.~Bickson, and C.~Guestrin.
\newblock {{P}ower{G}raph: {D}istributed Graph-Parallel Computation on Natural
  Graphs}.
\newblock In {\em Presented as part of the 10th {USENIX} Symposium on Operating
  Systems Design and Implementation ({OSDI} 12)}, pages 17--30. {USENIX}, 2012.

\bibitem[GNR14]{gemsa2014dynamiclabel}
A.~Gemsa, M.~N\"ollenburg, and I.~Rutter.
\newblock Evaluation of labeling strategies for rotating maps.
\newblock In {\em Experimental Algorithms (SEA'14)}, volume 8504 of {\em
  Lecture Notes in Computer Science}, pages 235--246. Springer, 2014.

\bibitem[HC14]{Heintz:2014}
B.~Heintz and A.~Chandra.
\newblock Beyond graphs: Toward scalable hypergraph analysis systems.
\newblock {\em SIGMETRICS Perform. Eval. Rev.}, 41(4):94--97, April 2014.

\bibitem[HHM{\etalchar{+}}]{kahyparWebsite}
V.~Henne, T.~Heuer, H.~Meyerhenke, P.~Sanders, S.~Schlag, and C.~Schulz.
\newblock {KaHyPar -- Karlsruhe Hypergraph Partitioning}.
\newblock {\url{http://kahypar.org/}}.

\bibitem[HK00]{HendricksonK00}
B.~Hendrickson and T.~G. Kolda.
\newblock {Graph Partitioning Models for Parallel Computing}.
\newblock {\em Parallel Computing}, 26(12):1519--1534, 2000.

\bibitem[HNS]{viecutWebsite}
M.~Henzinger, A.~Noe, and C.~Schulz.
\newblock {VieCut -- Vienna Minimum Cuts}.
\newblock {\url{http://viecut.taa.univie.ac.at/}}.

\bibitem[HO92]{hao1992faster}
J.~Hao and J.~B. Orlin.
\newblock A faster algorithm for finding the minimum cut in a graph.
\newblock In {\em Proceedings of the 3rd Annual ACM-SIAM Symposium on Discrete
  Algorithms}, pages 165--174. Society for Industrial and Applied Mathematics,
  1992.

\bibitem[HRW17]{henzinger2017local}
M.~Henzinger, S.~Rao, and D.~Wang.
\newblock Local flow partitioning for faster edge connectivity.
\newblock In {\em Proceedings of the 28th Annual ACM-SIAM Symposium on Discrete
  Algorithms}, pages 1919--1938. SIAM, 2017.

\bibitem[HS17]{hs2017sea}
T.~Heuer and S.~Schlag.
\newblock {Improving Coarsening Schemes for Hypergraph Partitioning by
  Exploiting Community Structure}.
\newblock In {\em {16th International Symposium on Experimental Algorithms,
  (SEA)}}, page 21:1–21:19, 2017.

\bibitem[KAKS99]{hMetisRB}
G.~Karypis, R.~Aggarwal, V.~Kumar, and S.~Shekhar.
\newblock {Multilevel Hypergraph Partitioning: Applications in VLSI Domain}.
\newblock {\em {IEEE Transactions on Very Large Scale Integration VLSI
  Systems}}, 7(1):69--79, 1999.

\bibitem[Kar00]{karger2000minimum}
D.~R. Karger.
\newblock Minimum cuts in near-linear time.
\newblock {\em Journal of the ACM}, 47(1):46--76, 2000.

\bibitem[Kar01]{karger2001randomized}
D.~R. Karger.
\newblock A randomized fully polynomial time approximation scheme for the
  all-terminal network reliability problem.
\newblock {\em SIAM Review}, 43(3):499--522, 2001.

\bibitem[KHT09]{10.1371/journal.pcbi.1000385}
S.~Klamt, U.~Haus, and F.~Theis.
\newblock {Hypergraphs and Cellular Networks}.
\newblock {\em PLoS Comput Biol}, 5(5):e1000385, 05 2009.

\bibitem[KK96]{karypis1996parallel}
G.~Karypis and V.~Kumar.
\newblock {Parallel Multilevel $k$-way Partitioning Scheme for Irregular
  Graphs}.
\newblock In {\em Proceedings of the ACM/IEEE Conference on Supercomputing'96},
  1996.

\bibitem[KR13]{Kirmani:2013:SPG:2503210.2503280}
S.~Kirmani and P.~Raghavan.
\newblock Scalable parallel graph partitioning.
\newblock In {\em Proceedings of of the International Conference on High
  Performance Computing, Networking, Storage and Analysis}, SC '13, pages
  51:1--51:10. ACM, 2013.

\bibitem[Kri84]{krishnamurthy1984improved}
B.~Krishnamurthy.
\newblock An improved min-cut algorithm for partitioning {VLSI} networks.
\newblock {\em IEEE Transactions on Computers}, 33(5):438--446, 1984.

\bibitem[KS96]{karger1996new}
D.~R. Karger and C.~Stein.
\newblock A new approach to the minimum cut problem.
\newblock {\em Journal of the ACM}, 43(4):601--640, 1996.

\bibitem[LCH17]{li2017efficient}
Y.~Li, S.~Cai, and W.~Hou.
\newblock An efficient local search algorithm for minimum weighted vertex cover
  on massive graphs.
\newblock In {\em Asia-Pacific Conference on Simulated Evolution and Learning
  (SEAL 2017)}, volume 10593 of {\em Lecture Notes in Computer Science}, pages
  145--157. 2017.

\bibitem[Lei80]{leiserson1980area}
C.~E. Leiserson.
\newblock {Area-Efficient Graph Layouts}.
\newblock In {\em 21st Symposium on Foundations of Computer Science}, pages
  270--281. IEEE, 1980.

\bibitem[LFK09]{lancichinetti2009detecting}
A.~Lancichinetti, S.~Fortunato, and J.~Kert{\'e}sz.
\newblock Detecting the overlapping and hierarchical community structure in
  complex networks.
\newblock {\em New Journal of Physics}, 11(3):033015, 2009.

\bibitem[LGH{\etalchar{+}}17]{li2017spac}
L.~Li, R.~Geda, A.~B. Hayes, Y.~Chen, P.~Chaudhari, E.~Z. Zhang, and
  M.~Szegedy.
\newblock A simple yet effective balanced edge partition model for parallel
  computing.
\newblock {\em SIGMETRICS Perform. Eval. Rev.}, 45(1):6--6, June 2017.

\bibitem[LSS{\etalchar{+}}]{kamisWebsite}
S.~Lamm, P.~Sanders, C.~Schulz, D.~Strash, and R.~F. Werneck.
\newblock {KaMIS -- Karlsruhe Maximum Independent Sets Homepage}.
\newblock {\url{http://algo2.iti.kit.edu/kamis/}}.

\bibitem[LT80]{lipton1980applications}
R.~J. Lipton and R.~E. Tarjan.
\newblock {Applications of a Planar Separator Theorem}.
\newblock {\em SIAM Journal On Computing}, 9(3):615--627, 1980.

\bibitem[Mat93]{matula1993linear}
D.~W. Matula.
\newblock A linear time $2+\varepsilon$ approximation algorithm for edge
  connectivity.
\newblock In {\em Proceedings of the 4th annual ACM-SIAM Symposium on Discrete
  Algorithms}, pages 500--504. SIAM, 1993.

\bibitem[MNS]{kadrawWebsite}
H.~Meyerhenke, M.~N\"ollenburg, and C.~Schulz.
\newblock {KaDraw -- Karlsruhe Graph Drawing Homepage}.
\newblock {\url{http://algo2.iti.kit.edu/kadraw/}}.

\bibitem[MWM15]{mccune-tlav-2015}
R.~R. McCune, T.~Weninger, and G.~Madey.
\newblock {Thinking Like a Vertex: A Survey of Vertex-Centric Frameworks for
  Large-Scale Distributed Graph Processing}.
\newblock {\em ACM Computing Survey}, 48(2):25:1--25:39, October 2015.

\bibitem[NI92]{nagamochi1992computing}
H.~Nagamochi and T.~Ibaraki.
\newblock Computing edge-connectivity in multigraphs and capacitated graphs.
\newblock {\em SIAM Journal on Discrete Mathematics}, 5(1):54--66, 1992.

\bibitem[NOI94]{nagamochi1994implementing}
H.~Nagamochi, T.~Ono, and T.~Ibaraki.
\newblock Implementing an efficient minimum capacity cut algorithm.
\newblock {\em Mathematical Programming}, 67(1):325--341, 1994.

\bibitem[NPS18]{hybrid-ils-2018}
B.~Nogueira, R.~G.~S. Pinheiro, and A.~Subramanian.
\newblock A hybrid iterated local search heuristic for the maximum weight
  independent set problem.
\newblock {\em Optimization Letters}, 12(3):567--583, 2018.

\bibitem[PBM10]{Patwary:2010}
M.~M.~A. Patwary, R.~H. Bisseling, and F.~Manne.
\newblock Parallel greedy graph matching using an edge partitioning approach.
\newblock In {\em Fourth International Workshop on High-level Parallel
  Programming and Applications}, HLPP '10, pages 45--54, 2010.

\bibitem[PM07]{Papa2007}
D.~A. Papa and I.~L. Markov.
\newblock {Hypergraph Partitioning and Clustering}.
\newblock In {\em {Handbook of Approximation Algorithms and Metaheuristics.}}
  2007.

\bibitem[PR90]{padberg1990efficient}
M.~Padberg and G.~Rinaldi.
\newblock An efficient algorithm for the minimum capacity cut problem.
\newblock {\em Mathematical Programming}, 47(1):19--36, 1990.

\bibitem[PR91]{padberg1991branch}
M.~Padberg and G.~Rinaldi.
\newblock A branch-and-cut algorithm for the resolution of large-scale
  symmetric traveling salesman problems.
\newblock {\em SIAM Review}, 33(1):60--100, 1991.

\bibitem[Pul09]{pullan-2009}
W.~Pullan.
\newblock Optimisation of unweighted/weighted maximum independent sets and
  minimum vertex covers.
\newblock {\em Discrete Optimization}, 6(2):214--219, 2009.

\bibitem[RAK07a]{labelpropagationclustering}
U.~N. Raghavan, R.~Albert, and S.~Kumara.
\newblock {Near Linear Time Algorithm to Detect Community Structures in
  Large-Scale Networks}.
\newblock {\em Physical Review E}, 76(3), 2007.

\bibitem[RAK07b]{raghavan2007near}
U.~N. Raghavan, R.~Albert, and S.~Kumara.
\newblock Near linear time algorithm to detect community structures in
  large-scale networks.
\newblock {\em Physical Review E}, 76(3):036106, 2007.

\bibitem[RC87]{ramanathan1987counting}
A.~Ramanathan and C.~J. Colbourn.
\newblock Counting almost minimum cutsets with reliability applications.
\newblock {\em Mathematical Programming}, 39(3):253--261, 1987.

\bibitem[RPGH14]{fatemeh2014jabejavc}
F.~Rahimian, A.~H. Payberah, S.~Girdzijauskas, and S.~Haridi.
\newblock Distributed vertex-cut partitioning.
\newblock In {\em Distributed Applications and Interoperable Systems}, pages
  186--200, Berlin, Heidelberg, 2014.

\bibitem[Shn96]{s-ehtdttiv-96}
B.~Shneiderman.
\newblock The eyes have it: a task by data type taxonomy for information
  visualizations.
\newblock In {\em Visual Languages (VL'96)}, pages 336--343. {IEEE}, 1996.

\bibitem[SL]{code}
C.~Stein and M.~Levine.
\newblock Minimum cut code.
\newblock \url{http://www.columbia.edu/~cs2035/code.html}.
\newblock Accessed: 2017-06-09.

\bibitem[SS]{kahipWebsite}
P.~Sanders and C.~Schulz.
\newblock {KaHIP -- Karlsruhe High Qualtity Partitioning Homepage}.
\newblock {\url{http://algo2.iti.kit.edu/kahip}}.

\bibitem[SS12]{kaffpaE}
P.~Sanders and C.~Schulz.
\newblock {Distributed Evolutionary Graph Partitioning}.
\newblock In {\em Proceedings of the 12th Workshop on Algorithm Engineering and
  Experimentation (ALENEX'12)}, pages 16--29, 2012.

\bibitem[SS13]{kabapeE}
P.~Sanders and C.~Schulz.
\newblock {Think Locally, Act Globally: Highly Balanced Graph Partitioning}.
\newblock In {\em Proceedings of the 12th International Symposium on
  Experimental Algorithms (SEA'12)}, Lecture Notes in Computer Science.
  Springer, 2013.

\bibitem[ST]{viemaWebsite}
C.~Schulz and J.~L. Träff.
\newblock {VieM -- Vienna Process Mapping and Sparse Quadratic Assignment}.
\newblock {\url{http://viem.taa.univie.ac.at/}}.

\bibitem[SWZ02]{schulz2002using}
F.~Schulz, D.~Wagner, and C.~Zaroliagis.
\newblock Using multi-level graphs for timetable information in railway
  systems.
\newblock In {\em Proceedings of Algorithm Engineering and Experiments
  (ALENEX)}, pages 43--59. Springer, 2002.

\bibitem[Tri06]{trifunovic2006parallel}
A.~Trifunovic.
\newblock {\em {Parallel Algorithms for Hypergraph Partitioning}}.
\newblock PhD thesis, University of London, 2006.

\bibitem[XKK16]{xu2016new}
H.~Xu, T.~S. Kumar, and S.~Koenig.
\newblock A new solver for the minimum weighted vertex cover problem.
\newblock In {\em International Conference on AI and OR Techniques in
  Constriant Programming for Combinatorial Optimization Problems}, pages
  392--405. Springer, 2016.

\end{thebibliography}


\begin{thebibliography}{100}


\bibitem{clustersnw}
Henning Meyerhenke, Peter Sanders and Christian Schulz.
\newblock {Partitioning Complex Networks via Size-constrained Clustering}.
\newblock In {\em Proceedings of the 13th Symposium on Experimental Algorithms (SEA)}, volume 8504 of Lecture Notes in Computer Science, pages 351--363. Springer, 2014.

\bibitem{externalcomplexnetworks}
Yaroslav Akhremtsev, Peter Sanders and Christian Schulz.
\newblock {(Semi-)External Algorithms for Graph Partitioning and Clustering}.
\newblock In {\em Proceedings of the 17th Workshop on Algorithm Engineering and Experimentation (ALENEX)}, pages 33--43. SIAM~2015.

\bibitem{parallelsnw}
Henning Meyerhenke, Peter Sanders and Christian Schulz.
\newblock {Parallel Graph Partitioning for Complex Networks}.
\newblock In {\em 29th IEEE International Parallel and Distributed Processing Symposium (IPDPS)}, 2015.

\bibitem{drawing}
Henning Meyerhenke, Martin Nöllenburg and Christian Schulz.
\newblock {Drawing Large Graphs by Multilevel Maxent-Stress Optimization}.
\newblock In {\em Proceedings of the 23rd International Symposium on
Graph Drawing \& Network Visualization (GD)}, volume 9411 of Lecture Notes in Computer Science, pages 30--43. Springer, 2015.

\bibitem{HGBisecNLevelTR}
Sebastian Schlag, Vitali Henne, Tobias Heuer, Henning Meyerhenke, Peter Sanders and Christian Schulz.
\newblock {$k$-way Hypergraph Partitioning via $n$-Level Recursive Bisection}.
\newblock In {\em Proceedings of the 18th Workshop on Algorithm Engineering and Experimentation (ALENEX)}, pages 53--67. SIAM 2016.

\bibitem{NS}
Peter Sanders and Christian Schulz.
\newblock {Advanced Multilevel Node Separator Algorithms}.
\newblock In {\em Proceedings of the 15th Symposium on Experimental Algorithms (SEA)}, volume 9685 of Lecture Notes in Computer Science, pages 294--309. Springer,~2016.

\bibitem{sharedmempart}
Yaroslav Akhremtsev, Peter Sanders and Christian Schulz.
\newblock {High-Quality Shared-Memory Graph Partitioning}.
\newblock In {\em Proceedings of the 24th International European Conference on Parallel Computing (Euro-Par)}, volume 11014 of LNCS, pages 659--671, 2018.

\bibitem{processmapping}
Christian Schulz and Jesper Larsson Träff.
\newblock {Better Process Mapping and Sparse Quadratic Assignment Problems}.
\newblock In {\em Proceedings of the 16th Symposium on Experimental Algorithms (SEA)}, volume 75 of LIPIcs, pages 4:1--4:15, 2017.

\bibitem{dagone}
Orlando Moreira, Merten Popp and Christian Schulz.
\newblock {Graph Partitioning with Acyclicity Constraints}.
\newblock In {\em Proceedings of the 16th Symposium on Experimental Algorithms (SEA)}, volume 75 of LIPIcs, pages 30:1--30:15, 2017.

\bibitem{clustersnwjv}
Henning Meyerhenke, Peter Sanders and Christian Schulz.
\newblock {Partitioning (Hierarchically Clustered) Complex Networks via Size-Constrained Graph Clustering}.
\newblock {\em ACM Journal of Heuristics}, Volume 22, Issue 5, pages 759--782, 2016.

\bibitem{parallelsnwjv}
Henning Meyerhenke, Peter Sanders and Christian Schulz.
\newblock {Parallel Graph Partitioning for Complex Networks}.
\newblock {\em IEEE Transactions on Parallel and Distributed Systems}, Volume 28, Issue 9, pages 2625--2638, 2017.


\bibitem{jvdrawing}
Henning Meyerhenke, Martin Nöllenburg and Christian Schulz.
\newblock {Drawing Large Graphs by Multilevel Maxent-Stress Optimization}.
\newblock {\em IEEE Transactions on Visualization and Computer Graphics}, Volume 24, Issue 5, pages 1814--1827. 2018. 




\bibitem{kerMIS}
Sebastian Lamm, Peter Sanders, Christian Schulz, Darren Strash and Renato F.\@ Werneck
\newblock {Finding Near-Optimal Independent Sets at Scale}.
\newblock In {\em Proceedings of the 18th Workshop on Algorithm Engineering and Experimentation (ALENEX)}, pages 138-150. SIAM 2016.

\bibitem{MISonthefly}
Jakob Dahlum, Sebastian Lamm, Peter Sanders, Christian Schulz, Darren Strash and Renato F.\@ Werneck.
\newblock {Accelerating Local Search for the Maximum Independent Set Problem}.
\newblock In {\em Proceedings of the 15th Symposium on Experimental Algorithms (SEA)},  volume 9685 of Lecture Notes in Computer Science, pages 118--133. Springer, 2016.

\bibitem{mincut}
Monika Henzinger, Alexander Noe, Christian Schulz and Darren Strash.
\newblock {Practical Minimum Cut Algorithms}.
\newblock In {\em Proceedings of the 20th Workshop on Algorithm Engineering and Experimentation (ALENEX)}, pages 48--61. SIAM 2018.

\bibitem{parallelmis}
Demian Hespe, Christian Schulz and Darren Strash.
\newblock {Scalable Kernelization for Maximum Independent Sets}.
\newblock In {\em Proceedings of the 20th Workshop on Algorithm Engineering and Experimentation (ALENEX)}, pages 223--237. SIAM 2018.

\bibitem{smexact}
Monika Henzinger, Alexander Noe and Christian Schulz.
\newblock {Shared-Memory Exact Minimum Cuts}.
\newblock In {\em 33rd IEEE International Parallel and Distributed Processing Symposium (IPDPS)}, \emph{to appear}, 2019. 

\bibitem{exactWMIS}
Sebastian Lamm, Christian Schulz, Darren Strash, Robert Williger and Huashuo Zhang.
\newblock {Exactly Solving the Maximum Weight Independent Set Problem on Large Real-World Graphs}.
\newblock In {\em Proceedings of the 21th Workshop on Algorithm Engineering and Experimentation (ALENEX)}, pages 144--158, SIAM, 2019. 

\bibitem{misatscalejv}
Christian Schulz, Darren Strash, Sebastian Lamm, Peter Sanders and Renato F.\@ Werneck.
\newblock {Finding Near-Optimal Independent Sets at Scale}.
\newblock {\em ACM Journal of Heuristics}, Volume 23, Issue 4, pages 207--229, 2017.

\bibitem{mincutjv}
Monika Henzinger, Alexander Noe, Christian Schulz and Darren Strash.
\newblock {Practical Minimum Cut Algorithms}.
\newblock Invited to special issue of {\em ACM Journal of Experimental Algorithms (ACM JEA) for ALENEX 2018}, Volume 23, pages 1.9:1--1.8:22,~2018.




\bibitem{graphgennew}
Daniel Funke, Sebastian Lamm, Peter Sanders, Christian Schulz, Darren Strash and Moritz von Looz.
\newblock {Communication-free Massively Distributed Graph Generation}.
\newblock In {\em 32nd IEEE International Parallel and Distributed Processing Symposium (IPDPS)}, pages 336--347, 2018. \textbf{Best Paper Award}. 


\bibitem{scalableedgepartitioning}
Sebastian Schlag, Christian Schulz, Daniel Seemaier and  Darren Strash.
\newblock {Scalable Edge Partitioning}.
\newblock In {\em Proceedings of the 21th Workshop on Algorithm Engineering and Experimentation (ALENEX)}, pages 211--2225, SIAM, 2019.

\bibitem{bagenjv}
Peter Sanders and Christian Schulz.
\newblock {Scalable Generation of Scale-free Graphs}.
\newblock {\em Information Processing Letters}. Volume 116, Article No. 7, pages 489--491, 2016.

\bibitem{jvgraphgennew}
Daniel Funke, Sebastian Lamm, Ulrich Meyer, Peter Sanders, Christian Schulz, Darren Strash and Moritz von Looz.
\newblock {Communication-free Massively Distributed Graph Generation}.
\newblock Invited to special issue of {\em Journal of Parallel and Distributed Computing for IPDPS'18}, to appear,~2019.

\bibitem{evoIS}
Sebastian Lamm, Peter Sanders and Christian Schulz.
\newblock {Graph Partitioning for Independent Sets}.
\newblock In {\em Proceedings of the 14th Symposium on Experimental Algorithms (SEA)}, volume 8504 of Lecture Notes in Computer Science, pages 68--81. Springer, 2015.

\bibitem{areaDesignGIS}
Nitin Ahuja, Matthias Bender, Peter Sanders, Christian Schulz and Andreas Wagner.
\newblock {Incorporating Road Networks into Territory Design}.
\newblock In {\em Proceedings of the 23rd International Conference on
Advances in Geographic Information Systems (GIS)}. ACM Press,~2015.

\bibitem{evoNS}
Peter Sanders, Christian Schulz, Darren Strash and Robert Williger.
\newblock {Distributed Evolutionary $k$-way Node Separators}.
\newblock In {\em Proceedings of the Genetic and Evolutionary Computation Conference (GECCO)}, pages 345--252, 2017. \textbf{Best Paper Nominee}.

\bibitem{evoHG}
Robin Andre, Sebastian Schlag and Christian Schulz.
\newblock {Memetic Multilevel Hypergraph Partitioning}.
\newblock In {\em Proceedings of the Genetic and Evolutionary Computation Conference (GECCO)}, pages 347--354, ACM, 2018. 

\bibitem{evoDAG}
Orlando Moreira, Merten Popp and Christian Schulz.
\newblock {Evolutionary Multi-level Acyclic Graph Partitioning}.
\newblock In {\em Proceedings of the Genetic and Evolutionary Computation Conference (GECCO)}, pages 331--339, ACM, 2018.



\bibitem{clustering}
Sonja Biedermann, Monika Henzinger, Christian Schulz and Bernhard Schuster.
\newblock {Memetic Graph Clustering}.
\newblock In {\em Proceedings of the 17th Symposium on Experimental Algorithms (SEA)}, volume 103 of LIPIcs, pages 3:1--3:15, 2018. 






\end{thebibliography}
}

\end{document}